\documentclass{article}
\usepackage{siamltex}

\title{Robust Optimal Experimental Design Accounting for Sensor Failure}
\date{}
\author{
Rebekah White$^1$, Chandler Smith$^2$, Drew Kouri$^3$,  Jace Ritchie$^4$, \\ Wilkins Aquino$^5$, and Timothy Walsh$^6$\\ 
\\ 
 $^5$Department of Mechanical Engineering and Materials Science \\
    Duke University, Durham, NC, 27708 \\ \\ 
 $^{1-4, 6}$Sandia National Laboratories \\
	       Albuquerque, NM, 87112 \\
}

\begin{document}

\maketitle

\begin{abstract}
    Optimal experimental design provides a way of determining \textit{a-priori} the best locations at which to place accelerometers in vibrations analysis experiments. 
    However, in practice, sensors often fail during experimentation due  high mechanical accelerations.
    There have been limited works exploring the use of robust OED in the context of vibrations analysis, where design spaces (i.e. candidate sensor locations and orientations) are high-dimensional and the finite-element models are expensive to compute.  
    Therefore, this work considers the application of more general robust OED formulations to such a structural dynamics problem. 
    We employ a relaxation-based approach that enables the use of efficient gradient-based optimization.
    Furthermore, we leverage a binary-inducing penalty during optimization to provide a binary sensor design as an alternative to leveraging post-optimization rounding heuristics. 
    We consider performance metrics based on the log-determinant of the parameter covariance as well those based on parameter and prediction mean-squared errors.
    We find that although robust and classical designs are similar for the structural dynamics problem of interest, robust designs outperform classical designs on average over relevant failure scenarios of interest. 
\end{abstract}

\section{Introduction}\label{sec:intro}
Modeling and simulation are essential for predicting structural responses in a wide range of engineering applications, including aerospace, civil, and mechanical systems. 
The underlying structural dynamics models used to predict responses are often comprised of unknown parameters (e.g. material properties, damping coefficients, or external forces) that must be calibrated using experimental data.
Because experimental data are often limited due to the high cost and logistical challenges associated with performing physical tests, it is crucial to predict \textit{a-priori} what experiments will yield the most informative data.

Optimal experimental design (OED) provides a powerful approach for determining the ``best'' experiments to perform given a budget.
In the context of calibration (parameter estimation) of structural dynamics models, OED has widely been used to predict the sensor locations and orientations that would be most informative; see~\cite{ostachowicz2019,tan2020,hassani2023} for comprehensive reviews.
Such formulations characterize \textit{informative data} via scalar-valued functions of the associated covariance matrix (or its inverse, the Fisher information matrix). While many criteria exist, $D$-optimality is a popular choice used to minimize the determinant of the parameter covariance~\cite{fedorov1972}.
Despite its utility, classical OED formulations can underperform when applied to real-world engineering problems.
For example, in vibrations analysis, the high mechanical accelerations can cause physical detachment of sensors, known as \textit{sensor dropout}, while amplitude measurements exceeding the dynamic range of the sensor can distort the data, known as \textit{sensor clipping}.
Such sensor failures, however, are not accounted for in the standard OED formulations.

While there exist robust (to sensor failure) OED formulations in literature (see \cite{andrews_robustness_1979, imhof2002optimal, imhof2004optimal, smucker_robustness_2017, limmun_generating_2023, hanson_exploring_nodate}), there have been limited works exploring robust strategies in the context of vibration testing.
The work of~\cite{An2022} constructs a robust OED objective that penalizes both information loss as well as clustering of sensors (configuring multiple sensors in a local region). To account for additional model uncertainties, the information loss is computed as the mean and variance of the $D$-optimal criteria over the sampled uncertainties. The aim is then to maximize the worst-case performance over failure scenarios.
To deal with the computational intractability of the resulting optimization problem,~\cite{An2022} leverages a reduced candidate sensor location space and Gaussian process surrogates of the parameter-to-observable map; such choices then enable the use of evolutionary optimization algorithms.
Note however, such an approach would not be amenable in general to high-dimensional design spaces and expensive OED objective function evaluations.
The work of~\cite{Yang2023} leverages a Bayesian approach where optimality is defined in terms of Bayes' risk. 
Here, the partial failure of sensors is considered though a sensor reliability risk function; however, total sensor failure is not addressed. 
To mitigate the computational costs of such an approach, approximations of the criterion as well as surrogates of the parameter-to-observable space are leveraged.

Although existing works address important questions regarding the robustness of sensor placement strategies for structural dynamics applications, these approaches are not generally applicable to high-dimensional design problems with computationally expensive forward models where the aim is to guard against complete sensor failure over multiple failure scenarios or with specified probabilities of failure.
As such, this work considers the application of more general robust OED strategies (see \cite{andrews_robustness_1979, imhof2002optimal, imhof2004optimal, smucker_robustness_2017}) to a three-dimensional structural dynamics vibration testing exemplar. 
We leverage asymptotic arguments to construct OED formulations similar to~\cite{imhof2002optimal,imhof2004optimal} that are amenable to scenarios where the probabilities of sensor failure are known or can be predicted. 
Additionally, we formulate alternative OED objectives that sample the failure scenarios of interest when failure probabilities cannot be predicted. 
While this formulation is similar to~\cite{smucker_robustness_2017, andrews_robustness_1976}, we leverage it to assess binary failure scenarios (i.e. a sensor fails or does not) rather than scenarios with fixed sensor failure probabilities.

To combat the computational complexity introduced by binary optimization (i.e. determining optimal sensor placements) over a high-dimensional design space (i.e. a large number of candidate locations), we leverage a \textit{relaxation} approach that is popular in the context of classical OED, which leverages fractional weighting of important sensors (rather than binary weights). 
Such an approach enables the use of efficient gradient-based computation with guaranteed convergence~\cite{Boyd2009}.
To produce binary designs, we leverage a binary-inducing penalty, known as the double-well penalty, which to our knowledge is novel in the context of OED. 
When characterizing performance of the optimal designs, we assess not only the optimization criteria but additionally explore performance in terms of parameter and prediction mean-squared errors, as well as performance with respect to a variety of failure scenarios, including those accounting for in the robust formulation as well as more extensive sensor failures.
Ultimately, we find that for this problem, classical and robust designs may share many optimal locations and perform similarly in terms of the relevant optimization criteria. 
However, the robust designs outperform classical on average when considering relevant failure scenarios.

The main contributions of this work are summarized as follows:
\begin{itemize}

    \item Demonstrate robust OED formulations in the context of structural dynamics problems, where two formulations are considered depending upon whether the probabilities of sensor failure can be predicted or not.
    

    \item Expand the postprocessing of optimal designs to consider additional metrics, such as parameter and prediction mean-squared errors, as well as evaluating performance over failure scenarios of interest. 

    \item Implement a relaxed approach to enable computationally efficient gradient-based optimization schemes that make robust OED tractable in the context of structural dynamics applications.

    \item Introduction of a novel \textit{double-well penalty} that can be used within optimization to induce sparse binary designs while avoiding full binary optimization.

    \end{itemize}

An outline of this paper is as follows. In~\Cref{sec:methods} we first introduce the model problem of interest (\Cref{sec:model_prob}), followed by the inverse problem formulation (\Cref{sec:inv_prob}) that allows us to calibrate the model.
In~\Cref{sec:oed}, we introduce the standard and robust OED formulations, including the double-well penalty (\Cref{sec:double_well}). 
In~\Cref{sec:post_proc}, we define the performance metrics of interest and discuss how designs are evaluated based on the practical interpretation of the OED optimization results. 
Computational results for the model problem are provided in~\Cref{sec:results}, with final conclusions drawn in~\Cref{sec:conclusions}

\section{Background and methodology}\label{sec:methods}

Here, we describe the optimal experimental design (OED) formulations needed to determine sensor placement strategies that are robust to failure or clipping behavior. 
While the framework presented generalizes across application problems of interest, for concreteness, we present the formulations relevant to the computational results that follow in~\Cref{sec:results}.  
We begin in~\Cref{sec:model_prob} by defining the forward model based on structural dynamics vibration response modeling.
Following this, we present the inverse problem in~\Cref{sec:inv_prob}, which aims to inform the uncertain point loads from acceleration measurements.
Lastly, in~\Cref{sec:oed}, we provide the experimental design formulations that determine the optimal accelerometer (sensor) placement strategies.
Here, we present both the standard and robust formulations. 
Additionally, we discuss aspects of the optimization and subsequent postprocessing of the optimal designs relevant to practical implementation in~\Cref{sec:post_proc}.

\subsection{Model problem}\label{sec:model_prob}

We first present the relevant theory underpinning  structural vibration analysis. 
Then we present the application problem considered in the computational results, noted as the \textit{wedding cake} problem.
To model structural vibrations of a linear, time-invariant system under time-harmonic excitation, we leverage the spatially discretized Helmholtz equation to model displacement ${\bm u}(\omega)$ as a function of frequency $\omega \in \mathbb{R}$ according to
\begin{eqnarray}\label{eq:Helmholtz}
    ({\bm K} - j\omega {\bm C} + \omega^2{\bm M}){\bm u}(\omega) =  {\bm p}(\omega),
\end{eqnarray}
where ${\bm p}(\omega)$ are external forces (loads), ${\bm M} \in \mathbb{R}^{N \times N}$ is the mass matrix, ${\bm C} \in \mathbb{R}^{N \times N}$ is the damping matrix, and ${\bm K} \in \mathbb{R}^{N \times N}$ is the stiffness matrix.
Note that the degrees of freedom (DoFs) $N = N_x d$ is comprised of the number of nodes in the spatial discretization $N_x$ and the degrees of freedom associated with each node, here $d=3$ (corresponding to $x$, $y$, and $z$ components).
See \cite{aarset2025global} for more details on structural dynamics modeling.
%
%
We can then compute the accelerations $\ddot{\bm u}(\omega)$ at every node of the spatial discretization as 
\begin{eqnarray}\nonumber
    \ddot{\bm u}(\omega) &=& -\omega^2 {\bm u}(\omega) \\\nonumber
    &=& -\omega^2 ({\bm K} - j\omega {\bm C} + \omega^2{\bm M})^{-1}{\bm p}(\omega) \\\label{eq:accel}
    &=& -\omega^2{\bm H}(\omega){\bm p}(\omega).
\end{eqnarray}
 
We then map the mathematical equation defined in~\eqref{eq:accel} to the observable quantities of interest as follows.
First, let us define the fixed frequency of interest as $\omega_0$, and let $\mathcal{Q} \in \mathbb{R}^{\datadim \times N}$ as an observation operator that picks off the solution (computed at each DoF) for locations and orientations along which we can observe acceleration.
Thus, the observed part of the response is given as 
\begin{eqnarray}\label{eq:obs_resp}
     {\mathcal{Q}}\ddot{\bm u}(\omega_0) 
    &=& -\omega^2 {\mathcal{Q}} {\bm H}(\omega_0){\bm p}(\omega_0).
\end{eqnarray}
Next, we define the the parameters of interest $\param = [\theta_1, \dots, \theta_{\paramdim}]^{\top} \in \Theta \subset \mathbb{R}^{\paramdim}$ as the nonzero loads corresponding to the frequency of interest $\omega_0$, which can be computed as
\begin{eqnarray}\label{eq:param}
    {\bm p}(\omega_0) &=& \mathcal{P}\param, 
\end{eqnarray}
where $\mathcal{P} \in \mathbb{R}^{N \times \paramdim}$ is an operator mapping the nonzero loads to the full discretized load vector ${\bm p}(\omega_0)$.
Substituting~\eqref{eq:param} into~\eqref{eq:obs_resp}, provides the parameter-to-observable map ${\bm f}({\param}): \Theta \subset \mathbb{R}^{\paramdim} \to \mathbb{R}^{\datadim}$:
\begin{eqnarray}\label{eq:forward_mod}
    {\bm f}(\param) &=& \transmat \param,
\end{eqnarray}
where $\transmat = -\omega^2{\mathcal{Q}} {\bm H}(\omega_0)\mathcal{P} \in \mathbb{R}^{\datadim \times \paramdim}$ is referred to as the frequency response function (FRF). 
We refer to the parameter-to-observable map henceforth as the \textit{model} for simplicity.

\begin{figure}[ht]
\centering
\includegraphics[width=3.5in]{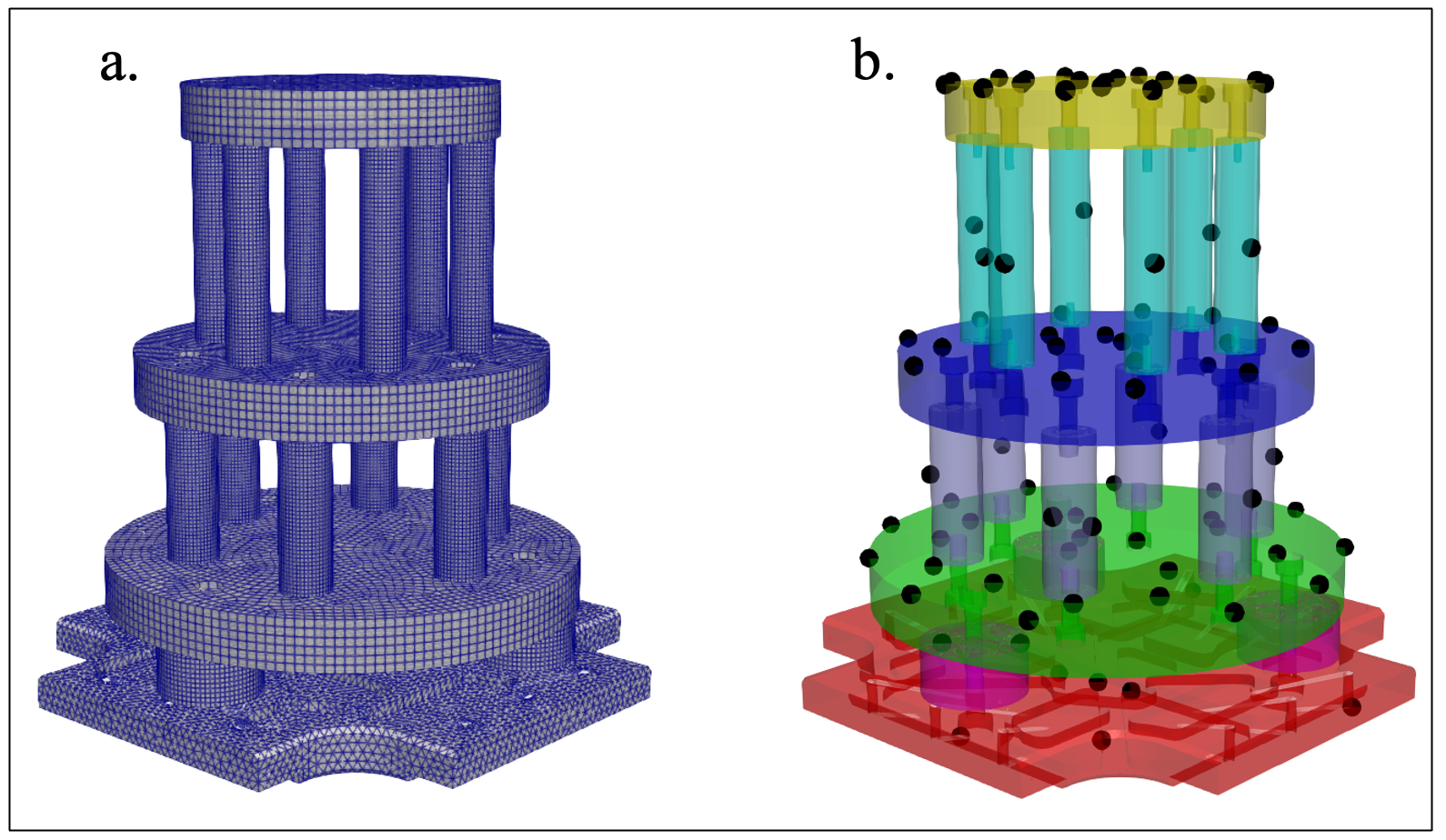}
\caption{ a) Finite element model of the wedding cake and 
b) candidate sensor locations superimposed over the model. Note that the horizontal levels of the structure are labeled as: base (red), level 1 (green), level 2 (blue), and level 3 (yellow).
Uniaxial sensors can be oriented in one of three degrees of freedom.}
\label{fig:candidate_sensor_fem}
\end{figure}

The application problem of interest is known as the \textit{wedding cake} problem and is depicted in~\Cref{fig:candidate_sensor_fem}.
We solve the discretized Helmholtz equation for this structure using Sandia's Sierra-SD finite element modeling (FEM) software~\cite{SierraSD}. 
In this example, there are $\datadim = 267$ candidate (acceleration) sensor degrees of freedom that can be observed, the locations of which are shown in~\Cref{fig:candidate_sensor_fem}b. 

\subsection{Inverse problem formulation}\label{sec:inv_prob}

The aim of the inverse problem is to leverage measurement data $\data \in \mathcal{Y} \subset \mathbb{R}^{\datadim}$ to inform the uncertain nonzero initial load forces $\param = [\theta_1, \dots, \theta_{\paramdim}]^{\top} \in \Theta = \mathbb{R}^{\paramdim}$.
%
First, we define the statistical model that represents the data collection process:
\begin{eqnarray}\label{eq:stat_mod}
    \data = \bm f(\param) + {\bm \epsilon} 
    = \transmat \param + {\bm \epsilon}, 
\end{eqnarray}
where we assume the covariance is of the form ${\bm \Gamma}_{\text{noise}} = \sigma^2{\bm I}$ and $\mathbb{E}[{\bm\epsilon}]={\bm 0}$.
We define the experimental design $\des \in \mathbb{R}^{\desdim}$ as the optimal degrees of freedom (DoFs) at which to collect data $\data$ given a maximum budget $b$ \footnote{Without loss of generality, we assume that every observable DoF is a candidate DoF. However, one could also restrict the design to a subset of the observable DoFs.}.
Note that for simplicity, we will refer to the candidate DoFs as \textit{candidate sensors} and the optimal design as the \textit{optimal sensor placement strategy}, while always referring to both location and orientation
Since for every DoF $i$ one can either collect data ($w_i = 1$) or not ($w_i = 0$), we mathematically represent the design as binary, i.e. $\des = [w_1, \dots w_{\desdim}]^\top$, for $w_i \in \{0, 1\}$.

For binary sensor placement problems, the experimental design enters the ordinary least squares (OLS) inverse problem~\cite{banks2009inverse}
as follows:
\begin{eqnarray}\label{eq:des_least_squares}
    = 
    \hat{\param} =
    \text{arg}\min_{\param \in \Theta} \sum_{i=1}^{\datadim} w_i\big( y_i - \transmat^{\top}_i{\param}\big)^2
    = \text{arg}\min_{\param \in \Theta} \| {\bm W} (\data - \transmat \param) \|_{2}^{2},
\end{eqnarray}
where $\transmat_i \in \mathbb{R}^{\paramdim}$ is the $i^{\text{th}}$ row of $\transmat$ and ${\bm W} = \text{diag}(w_1, \dots, w_{\desdim}) \in \mathbb{R}^{\desdim \times \desdim}$.
Due to linearity of the model, assumptions regarding the noise, and the fact that $w_i \in \{0, 1\}$,
the corresponding design-dependent covariance matrix is then given as
\begin{eqnarray}\nonumber
    {\bm C}(\des) 
    &=&  (\transmat^\top{\bm W}\transmat)^{-1}[\transmat^\top{\bm W}{\bm\Gamma}_{\text{noise}}{\bm W}\transmat](\transmat^\top{\bm W}\transmat)^{-1} \\\label{eq:weight_cov}
     &=&
    \sigma^2(\transmat^\top{\bm W}\transmat)^{-1}
\end{eqnarray}
Although we derive the covariance in~\eqref{eq:weight_cov} for the choice of statistical model and design variables of interest in this work~\eqref{eq:stat_mod}, one could analogously derive covariance matrices for more complex choices.

\subsection{Robust optimal experimental design formulation}\label{sec:oed}

Here, we present formulations of the robust OED problems considered in this work. 
We determine optimal designs by defining OED criteria $\Psi({\bm C}(\des))$ that provide scalar-valued representations of the uncertainty characterized by the covariance.
Popular choices include those based on the trace and determinant, which are respectively noted as $A$- and $D$-optimal design, as well as $I$-optimal design penalizing prediction uncertainty \cite{hanson_exploring_nodate, imhof2002optimal, lee2018optimal,limmun_generating_2023,smucker_robustness_2017}.
Henceforth, we will note the criterion specifically as $\Psi := \log \det({\bm C}(\cdot))$, as this is used in the computational results that follow.
The full binary OED problem then determines an optimal sensor placement strategy $\des^*$ by solving
\begin{eqnarray}\label{eq:full_opt_des}
    \des^* = \operatorname*{argmin}_{\des \in \{0, 1\}^{\desdim}} \log \det({\bm C}(\des)) \quad \text{subject to} \quad \sum_{i=1}^{\desdim} w_i c_i \leq b, 
\end{eqnarray}
where $c_i$ represents a cost associated with each candidate sensor.

Binary optimization of~\eqref{eq:full_opt_des} over the space of all possible designs---of dimension $\desdim \choose b$---is computationally intractable for high-dimensional design spaces (e.g. if $\desdim$ is large) and expensive forwards models (e.g. those governed by differential equations), and thus, cannot be afforded in practice.
Therefore, this work considers a popular \textit{relaxed approach}, wherein the weights can take on fractional values $w_i \in [0, 1]$, for $i = 1, \dots, \desdim$; 
this enables the use of efficient gradient-based optimization~\cite{Boyd2009}.
The relaxed OED problem then becomes 
\begin{eqnarray}\label{eq:class_opt_des}
    \des^* = \operatorname*{argmin}_{\des \in [0, 1]^{\desdim}} \log \det({\bm C}(\des)) + \gamma P(\des) \quad \text{subject to} \quad \sum_{i=1}^{\desdim} w_i c_i \leq b,
\end{eqnarray}
where $P(\des)$ is a penalty that induces binary structure, tuned by the hyperparameter $\gamma$.
The double-well penalty used in this work is further discussed in~\Cref{sec:double_well}.

The optimal design given by~\eqref{eq:class_opt_des}, however, does not account for the fact that in a real experiment, sensors may fail and consequently is not robust to such failures.
To understand how to formulate robust sensor placement strategies, first consider that failure can be modeled in the inverse problem as
\begin{eqnarray}\label{eq:do_least_squares}
\hat{\param} &=& \text{arg}\min_{\param \in \Theta} \sum_{i=1}^{\datadim} \xi_i w_i \big( y_i - \transmat^{\top}_i{\param}\big)^2,
\end{eqnarray}
where $\xi_i = 0$ indicates sensor failure at DoF $i$, otherwise $\xi_i = 1$.
Since OED aims to predict optimal designs \textit{a-priori}, and we cannot know in advance which sensors will fail in practice, we model this probabilistically using Bernoulli random variables
$\Xi_i \sim \text{Bernoulli}(1-q_i)$, which realize values ($\Xi_i=\xi_i$) as follows: 
\begin{eqnarray}\label{eq:bernoulli}
    \xi_i = 
    \begin{cases}
        1, \quad \text{with probability} \quad 1-q_i \\
        0, \quad \text{with probability} \quad q_i, \quad i = 1, \dots, \desdim.
    \end{cases}
\end{eqnarray}
Furthermore, $\mathbb{E}[\Xi_i] = 1 - q_i$.
Leveraging the central limit theorem for $M$-estimators and the assumption that $\Xi_i$ is independent of the noise $\epsilon$ as well as $\Xi_j$ for $i\neq j$, we can write the covariance estimator that incorporates dropout as 
\begin{eqnarray}\label{eq:do_cov}
    {\bm C}(\pmat \des) 
    &=& \sigma^2\left( \sum_{i=1}^{\datadim} (1-q_i)w_i \transmat_i \transmat^{\top}_i\right)^{-1}, 
    \quad \sigma^2 = \frac{1}{n-\paramdim} \sum_{i=1}^{\datadim} (1-q_i) w_i\big( y_i - \transmat^{\top}_i\hat{\param}\big)^2,
\end{eqnarray}
where $\pmat \in \mathbb{R}^{\desdim \times \desdim}$ is a diagonal matrix such that $\lpmat_{ii} = 1 - q_i$, for $q_i \in [0, 1]$, representing the probability of failure associated with DoF $i$.
Note that alternative formulations of OED criteria that account for specified probabilities of failure leverage non-asymptotic covariance structures.  
However, such formulations are significantly more computationally expensive, often requiring simplifying assumptions to maintain tractability;
see~\hyperref[sec:app]{Appendix} for a more detailed comparison.
Therefore, the robust optimal experimental design problem is 
\begin{eqnarray}\label{eq:rob_opt_des}
     \des^* = \operatorname*{argmin}_{\des \in [0, 1]^{\desdim}} \log \det\big({\bm C}(\pmat\des)\big) + \gamma P(\des) \quad \text{subject to} \quad \sum_{i=1}^{\desdim} w_i c_i \leq b.
\end{eqnarray}
Notice, the higher the probability of failure associated with a given sensor ($q_i \to 1$), 
the more sensor $i$ becomes downweighted ($(1-q_i)w_{i} \to 0$) due to the high likelihood of that sensor failing.
If one wants to consider the scenario where a specific DoF $i$ fails, this is analogous to specifying $q_i = 1$.

If precise knowledge regarding the probabilities of failure associated with each sensor is not known, one can sample realizations of probabilities of failure by constructing realizations of the associated matrices $\pmat_j$.
Here, the optimal design is determined according to 
\begin{eqnarray}\label{eq:mult_rob_opt_des}
    \des^* = \operatorname*{argmin}_{\des \in [0, 1]^{\desdim}} \frac{1}{\sendim} \sum_{j=1}^{\sendim} \log \det\big({\bm C}(\pmat_j \des)\big) + \gamma P(\des) \quad \text{subject to} \quad \sum_{i=1}^{\desdim} w_i c_i \leq b,
\end{eqnarray}
where $\sendim$ is the number of failure probability samples.
Note that average in~\eqref{eq:mult_rob_opt_des}, could be replaced with the maximum or conditional value at risk~\cite{kouri_risk-adapted_2022}, if one  wanted to penalize worst-case information loss rather than expected loss.

Without relying on sampling probabilities, the formulation given by~\eqref{eq:mult_rob_opt_des} can also be utilized to design sensor placement strategies that are robust to the failure of any number of sensors. 
To see this, consider that there are $\desdim$ candidate sensor locations, and one wants to be robust to one sensor failure.
Thus, for $\sendim =$ $\desdim \choose 1$, one constructs diagonal matrices $\{\pmat_j\}_{j=1}^{\sendim}$, such that $(\pmat_j)_{ii} = 0$ for $i=j$ and solves~\eqref{eq:mult_rob_opt_des}.
In doing so, we are directly considering the average performance with respect to each of the $\desdim$ sensors failing.
Note however, as the number of failure scenarios increases (e.g. $\sendim =$ $\desdim \choose b$, for $b=2, 3, \dots$) there is a significant increase the computational cost associated with solving~\eqref{eq:mult_rob_opt_des}, although
this may be justifiable in comparison to the cost of experimentation.

%
%

\subsubsection{Double-well penalty}\label{sec:double_well}

In relaxed formulations of OED problems, a penalty function $P(\des)$ can be used to enforce sparse designs. 
Well known choices include those based on $l_0$- or $l_1$-norms~\cite{alexanderian_optimal_2021}.
Here, we utilize a binary-inducing penalty that, to our knowledge, is novel in the context of OED.
This so-called \textit{double-well penalty} is defined as
\begin{equation}\label{eq:double_well}
    P(\des) = \sum_{i=1}^{\desdim} w_i (1 - w_i).
\end{equation}
We can intuitively understand the double well penalty by noting that the minima of~\eqref{eq:double_well} on the set $[0,1]^{n_y}$ occur at one and zero; hence when leveraged within optimization, fractional solutions tend to be penalized.

We note that even without an additive penalty $P(\des)$, the combination of OED criterion along with a budget constraint ensures that resulting designs will be sparse;
see~\cite{pukelsheim2006optimal} for more details. 
However, the double-well penalty is employed in~\Cref{sec:results} to demonstrate how one can leverage the relaxed optimization formulation in a way that drives the optimal design towards binary structures \textit{during} optimization. 
Such an approach is an alternative to leveraging rounding heuristics \textit{after} optimization to estimate a binary design.

Computationally, the addition of penalty terms in the OED formulation (see~\eqref{eq:class_opt_des}-\eqref{eq:mult_rob_opt_des}), introduces non-convexity into the optimization problem, allowing gradient-based methods to converge to suboptimal local minima.
To address this challenge, we compute the optimal design for varying penalty weights 
$\gamma$. 
We then select as the optimal design, the solution that best satisfies the budget constraint while also minimizing the OED criterion. 
Despite the heuristic nature of such an approach, we find it to be both efficient and effective in practice, yielding binary designs that balance the sensor budget constraints with the goal of increasing information gain in a way that is robust to sensor failure.

\subsection{Evaluating the performance of experimental designs}\label{sec:post_proc}

A straightforward approach to comparing the performance of classical versus robust optimal designs (determined via the OED objectives outlined in~\Cref{sec:oed}) is to compute the respective 
criteria, defined as: 
\begin{eqnarray}\label{eq:class_perf}
    \Psi_{\text{D-opt}}(\des) &=& \log \det\big({\bm C}(\des)\big), \\\label{eq:robust_perf}
    \Psi_{\text{robust}}(\des) &=& \frac{1}{\sendim} \sum_{j=1}^{\sendim} \log \det\big({\bm C}(\pmat_j\des)\big),
\end{eqnarray}
where $S$ is the number of failure scenarios considered in the optimization of the robust design. 
However, it is also informative to compare designs in terms of additional metrics, referred to as postprocessing metrics to indicate these were not leveraged in the optimization of the designs. 
In~\Cref{sec:mse,sec:fail_pp,sec:bern_pp} we present the metrics considered in the computational results and provide intuition for why such postprocessing analysis is useful in assessing performance.

\subsubsection{Evaluating performance based on parameter and prediction mean squared errors}\label{sec:mse}

The mean-squared errors (MSEs) associated with a design provide a measure of how accurate a corresponding parameter or prediction estimate will be from the true value.
Let $\param_0$ be the unknown ``true'' parameter, and let $\hat{\param}(\des)$ be the estimator computed from data $\data$ collected according to a design $\des$.
When not accounting for sensor failure, due to the linearity of the model along with the noise assumptions, the covariance~\eqref{eq:weight_cov} can be computed without relying on asymptotic assumptions and provides a direct measure of an estimators associated mean-squared error (MSE).
The parameter and prediction MSE are given respectively as
%
%
\begin{eqnarray}
    \text{MSE}\big(\hat{\param}(\des)\big) &=& 
    \mathbb{E}\left[ \|\hat{\param}(\des)-\param_0\|_{2}^2 \right]
    \\
    &=& \text{trace}\big({\bm C}(\des)\big),
\end{eqnarray}
and
\begin{eqnarray}
    \text{PMSE}\big(f(\hat{\param}(\des))\big) &=&
    \mathbb{E}\left[ \|  f(\hat{\param}(\des)) - f(\param_0) \|_2^2\right]
    \\ 
    &=&
    \mathbb{E}\left[ \|\transmat\hat{\param}(\des)-\transmat{\param}_0 \|_2^2\right]
    \\
    &=&
    \text{trace}\big(\transmat {\bm C}(\des) \transmat^{\top}\big).
\end{eqnarray}

%
When incorporating sensor dropout, however, one can no longer rely on the resulting covariance as a direct measure of MSE; 
this is because the covariance given in~\eqref{eq:do_cov} now relies on asymptotic approximation, meaning it is valid in the limit of infinite data, but may not reflect the true errors computed with finite data samples.
As a result, we evaluate the performance of dropout designs in terms of the \textit{empirical~MSEs}, i.e. those approximated from finite data.
Here, we leverage Monte Carlo (MC) sampling to approximate the integral given by the expectation. With a nominal parameter $\param_0$, we  sample the noise $\{\bm \epsilon^{(i)} \}_{i=1}^{\testdim} \sim \mathcal{N}({\bm 0}, {\bm \Gamma}_{\text{noise}})$ and generate realizations of data 
\begin{eqnarray}
    \data^{(i)} = \transmat \param_0 + {\bm \epsilon}^{(i)}, \quad i = 1, \dots, \testdim,
\end{eqnarray}
and compute the corresponding parameter estimates $\hat{\param}^{(i)}(\des)$.
The \textit{empirical (scalar) parameter MSE} is then given as as 
\begin{eqnarray}\label{eq:est_param_mse}
    \text{MSE}_{\text{empirical}}\big(\hat{\param}(\des)\big) &\approx& \frac{1}{\testdim} \sum_{i=1}^{\testdim} \| \hat{\param}^{(i)}(\des)- \param_0\|^2.
\end{eqnarray}
Analogously,
the scalar \textit{empirical (scalar) prediction MSE} is given as
\begin{eqnarray}\label{eq:pred_mse}
    \text{PMSE}_{\text{empirical}}\big(f\big(\hat{\param}(\des)\big)\big) &\approx& \frac{1}{\testdim} \sum_{i=1}^{\testdim} \| {\bm f}\big(\hat{\param}^{(i)}(\des)\big) - {\bm f}\big(\param_0\big) \|^2
    \\
    &=&
    \frac{1}{\testdim} \sum_{i=1}^{\testdim} \|\transmat\hat{\param}^{(i)}(\des) - \transmat \param_0 \|^2.
\end{eqnarray}
In the computational results that follow, we explore behavior as sensors fail by computing the average (over samples of $\param_0 \sim \pi(\param_0)$) empirical MSEs for each failure scenario.

\subsubsection{Evaluating performance over failure scenarios}\label{sec:pp_fail_scen}

The log-determinant and mean squared errors (parameter and prediction) can be utilized to evaluate design performance over failure scenarios of interest in a postprocessing context.
Note this differs from~\eqref{eq:robust_perf}), as we are not restricted to failure scenarios considering during optimization.
Here, we can define the performance criteria as 
\begin{eqnarray}\label{eq:multi_fail_post_proc}
     \phi^{(j)}(\des) &=& \log \det\big({\bm C}(\tilde{\pmat}_j\des)\big), \\
     \label{eq:multi_fail_post_proc_mse}
    \phi^{(j)}(\des) &=& \text{MSE}_{\text{empirical}}\big(\hat{\param}(\tilde{\pmat}_j\des)\big) \\
    \phi^{(j)}(\des) &=& \text{PMSE}_{\text{empirical}}\big(\hat{\param}(\tilde{\pmat}_j\des)\big), 
\end{eqnarray}
where the construction of $\tilde{\pmat}_j$ for $j=1, \dots, \tilde{S}$ emphasizes the ability to explore failure scenarios differing from the those considered during optimization.
Next, we discuss in detail how such $\tilde{\pmat}_j$ are constructed for the postprocessing analysis considered in~\Cref{sec:results}.

\vspace{0.4cm}
\noindent\textbf{Failure of one or more optimal sensors}
\vspace{0.1cm}

Once optimal designs have been determined, one can compare performance in terms of failure of the \textit{optimal} sensors, rather than the \textit{candidate} sensors as in~\eqref{eq:robust_perf}.
For example, if $n_{\text{opt}}$ represents the number of sensors in an optimal design. 
One may evaluate performance $\phi^{(j)}$ as one of the $n_{\text{opt}}$ sensors fail, i.e. for $j=1, \dots, \tilde{S} =$ $n_{\text{opt}} \choose 1$.
By considering the failure of more than one (up to the point where the inverse problem becomes ill-posed) optimal sensors, we can understand if designs optimized for \textit{one sensor failure} are robust to additional failures. 
Evaluating if designs have additional robustness in postprocessing is beneficial as directly optimizing for multiple sensor failures is often computationally prohibitive due to the combinatorial complexity. 
By evaluating the distribution over $j$, we can explore both the average and worst-case performance computed respectively as 
\begin{eqnarray}\label{eq:av_perf}
    \frac{1}{\tilde{S}}\sum_{j=1}^{\tilde{S}} \phi^{(j)}(\des) \\\label{eq:worst_case_perf}
    \max_{j} \phi^{(j)}(\des)
\end{eqnarray}
We note that 
evaluating~\eqref{eq:multi_fail_post_proc} provides an analogous comparison to evaluating $D$-efficiency~\cite{smucker_robustness_2017}.

\vspace{0.4cm}
\noindent \textbf{Failure scenarios sampled from sensor probabilities of failure}
\vspace{0.1cm}

When robust designs are computed by specifying probabilities of failure (PoFs), we can evaluate performance with respect to optimal sensors failing at rates specified by the PoFs.
That is, we simulate a series of possible experimental scenarios where sensors fail based on our beliefs about the likelihood of failure. 
We then compare how well the classical and robust designs perform over such failure scenarios.
Here, $\tilde{\pmat}_j$ is determined by sampling Bernoulli random variables $\xi^{(j)}_i \sim \text{Bernoulli}(1-q_i)$ for, $i = 1, \dots, \desdim$, where $j=1,\dots,n_{\text{samps}}$ refers to the number of simulated experiments one samples over.
Here, we evaluate average performance~\eqref{eq:av_perf} across all sampled failure scenarios, even those for which the optimal sensors are not affected; this provides both a notion of how likely failure of the optimal sensors is based on the PoFs as well as the impact of optimal sensor failures on performance.
This is a key difference from evaluating failure of one of more optimal sensors, where probabilistic arguments are not leveraged in the optimization problem and hence, the focus is only on performance when the optimal sensors fail.
Note that if a sampled $\tilde{\pmat}_j$ results in the inverse problem becoming ill-posed (i.e. too many optimal sensors are dropped), that element of the sum is set to $0$.
We note that sampling Bernoulli random variables and evaluating the mean $D$-optimal criterion is similar to what alternative robust OED formulations leverage when optimizing designs~\cite{smucker_robustness_2017}. 
See~\hyperref[sec:app]{Appendix} for further discussion and comparison of the approaches. 

\vspace{0.4cm}
\noindent \textbf{Failure scenarios corresponding to clipping behavior}
\vspace{0.1cm}

Sensor clipping occurs when the measured vibration exceeds the dynamic range of the accelerometer, leading to distortion or complete loss of data, which we model as sensor dropout. 
Here, we use model predictions corresponding to random force realizations to determine which sensors will be clipped and hence, how we construct $\tilde{\pmat}_j$. Here, $j =1, \dots, \clipdim$, where $\clipdim$ refers to the number of force realizations considered when predicting acceleration responses. 
For a given force realization $j$, 
if candidate accelerometer sensor $i$ is predicted to exceed the clipping threshold, it is noted as a dropout sample with $\{\tilde{\pmat}_j\}_{ii} = 0$.
We then evaluate performance $\phi^{(j)}$ with respect to all failure scenarios, even if such scenarios do not result in an optimal sensor failing. 
Doing so provides a sense of both how often the clipping behavior  impacts the optimal sensors and what that impact is in terms of performance.
Note here that the failure scenarios considered during optimization are exactly those considered in postprocessing ($\pmat_j = \tilde{\pmat}_j$), but the analysis is conducted in terms of a histogram of performances $\phi^{(j)}$ over the failure scenarios as well as computation of the mean performance given by~\eqref{eq:robust_perf}.
%


%

\section{Computational Results}\label{sec:results}

We apply the robust formulations given in~\Cref{sec:oed} to the wedding cake model described in~\Cref{sec:model_prob}.
We first determine designs that are robust to one sensor failure in~\Cref{sec:one_sensor_fails}. 
Then, in~\Cref{sec:comp_res_known_pof} we consider robust designs that are determined based on known or predicted probabilities of sensor failure, something that is common in many engineering scenarios.
Lastly, in~\Cref{sec:clipping} we demonstrate how dropout formulations can be used to address clipping behavior, wherein data is distorted or lost due to high-magnitude responses (e.g. accelerations), a common feature in real-world engineering applications.
In evaluating the performance of classical versus robust designs, we leverage the postprocessing criteria described in detail in~\Cref{sec:post_proc}.

\subsection{Robustness to one sensor failure}\label{sec:one_sensor_fails}

Here, we explore how the optimal experimental designs vary as we account for the possibility of any one sensor failing during an experiment. 
Since we do not have \textit{a-priori} knowledge about which sensors are most likely to fail, we leverage the OED objective function defined in~\eqref{eq:mult_rob_opt_des} where $S =$ $\desdim \choose 1$ to compute the robust experimental design.
We note that for the computational example defined in~\Cref{sec:model_prob}, when leveraging post-optimization rounding heuristics or the binary-inducing double-well penalty, the robust and classical optimal designs are the same. 
This indicates that for some problems and restrictions of the designs space (binary designs), the classical D-optimal design may be inherently robust to sensor failure and is in line with others' findings~\cite{smucker_robustness_2017}. 

\begin{figure}[h!]
    \centering
    \includegraphics[width=0.7\linewidth]{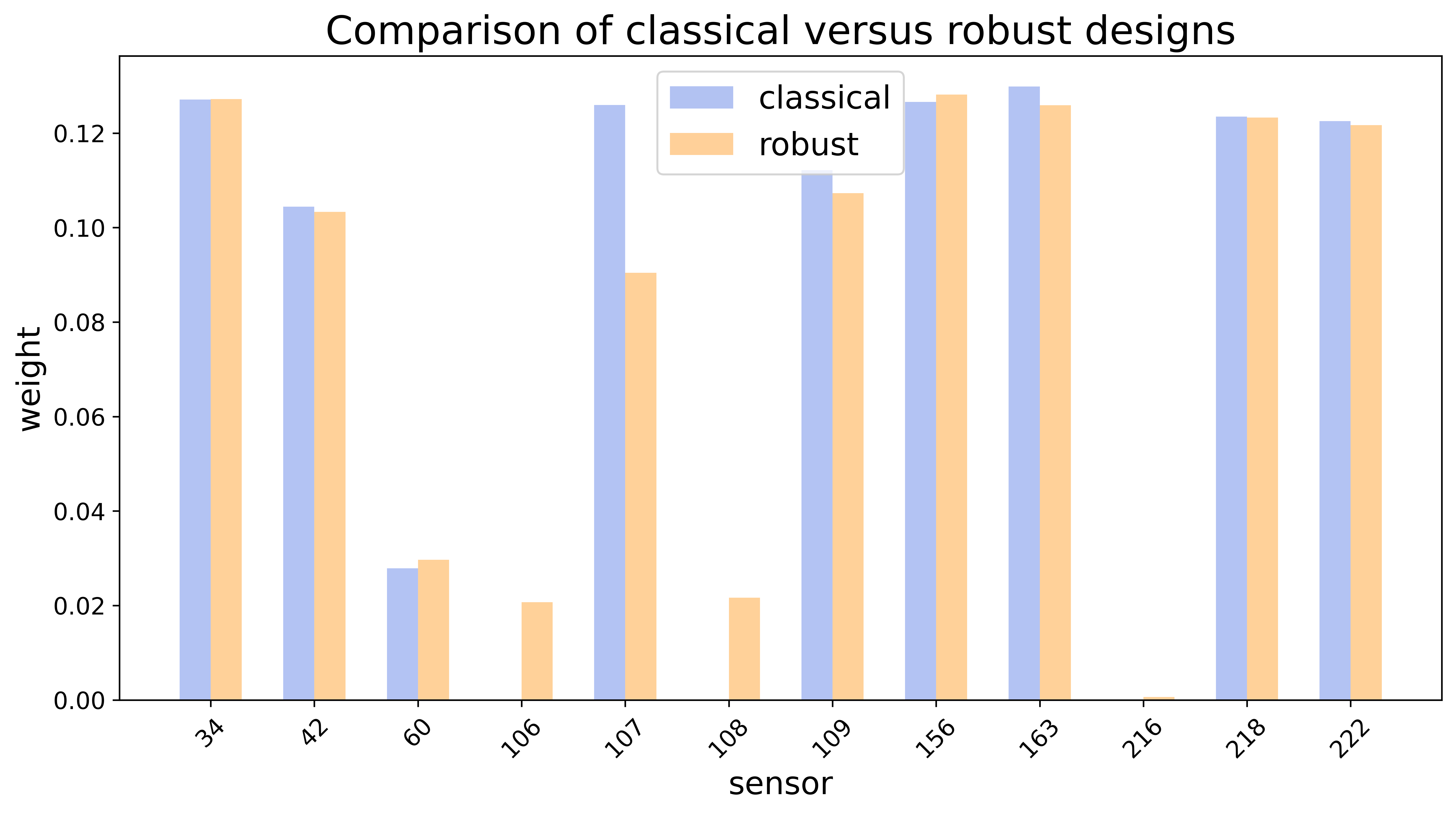}
    \caption{The fractional classical and robust optimal experimental designs determined by optimizing the OED objectives in~\eqref{eq:class_opt_des} and~\eqref{eq:mult_rob_opt_des}, respectively, where no binary-inducing penalty is applied, i.e. $\gamma = 0$.}
    \label{fig:AllDropOut_rob_v_class_frac_des}
\end{figure}

However, if one removes the binary design restriction and considers fractional designs resulting from solving the relaxed formulation of the optimization problem without a binary-inducing penalty ($\gamma=0$) or post-optimization rounding, advantages of robust designs over classical can be found. 
In this context, the fractional weights can be interpreted as frequencies of measuring at specific locations, when the experiments can be repeated. 
Although such a scenario is not the focus of the application highlighted in this work, we present the findings as they provide intuition for the robust formulations given in~\Cref{sec:oed}, are relevant for related problems of interest, and go beyond the analysis currently found in literature.
A comparison of the classical and robust designs corresponding to the OED objective functions given respecitively in~\eqref{eq:class_opt_des} and~\eqref{eq:mult_rob_opt_des} is given in~\Cref{fig:AllDropOut_rob_v_class_frac_des}.
Here, we notice that the two fractional designs differ, with the robust design having more nonzero weights, i.e. being less sparse, which intuitively, is advantageous if one suspects a sensor may fail.

\begin{figure}[h!]
    \centering
    \includegraphics[width=0.45\linewidth]{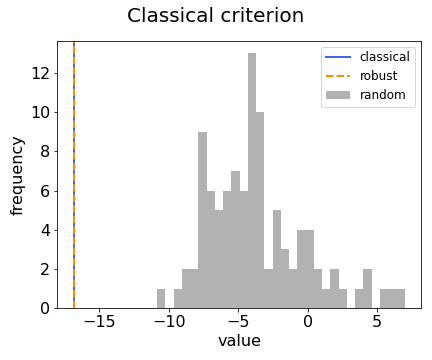} 
    \includegraphics[width=0.45\linewidth]{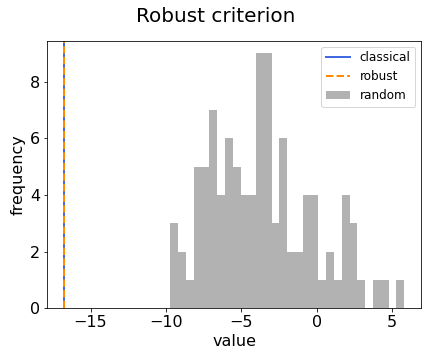}
    \caption{Comparing the classical and robust fractional optimal designs versus random  designs based on the (left) classical OED criterion in~\eqref{eq:class_perf} versus the (right) robust criterion in~\eqref{eq:robust_perf}.}
    \label{fig:AllDropOut_rob_v_class_frac_performance}
\end{figure}

We further compare the performance of the optimal designs versus random designs in terms of their respective criteria in~\Cref{fig:AllDropOut_rob_v_class_frac_performance}.
Note that when comparing fractional designs, we compute random designs as follows. 
Let $\tilde{N}$ represent the maximum number of nonzero weights in the robust and classical designs (for example, from~\Cref{fig:AllDropOut_rob_v_class_frac_des}, $\tilde{N} = 12$). 
Then, randomly choose $\tilde{N}$ of the $\desdim$ candidate sensor locations. 
To assign random fractional weights, we draw $\tilde{N}$ samples from a uniform distribution $\mathcal{U}([0, 1))$ and renormalize the samples.
From the left plot of~\Cref{fig:AllDropOut_rob_v_class_frac_performance}, we see that if no optimal sensors fail, the robust design performs as well as the classical design; this is a desired feature as one would not want a design that is robust to sensor failure to significantly underperform in the event a sensor did not fail. 
Similarly, when averaging performance over the set of all possible failure scenarios, i.e. evaluating the robust criterion, we see from the right plot of~\Cref{fig:AllDropOut_rob_v_class_frac_performance} that robust and classical designs have similar performance, both outperforming random designs.

\begin{figure}[h!]
    \centering
    \includegraphics[width=0.45\linewidth]{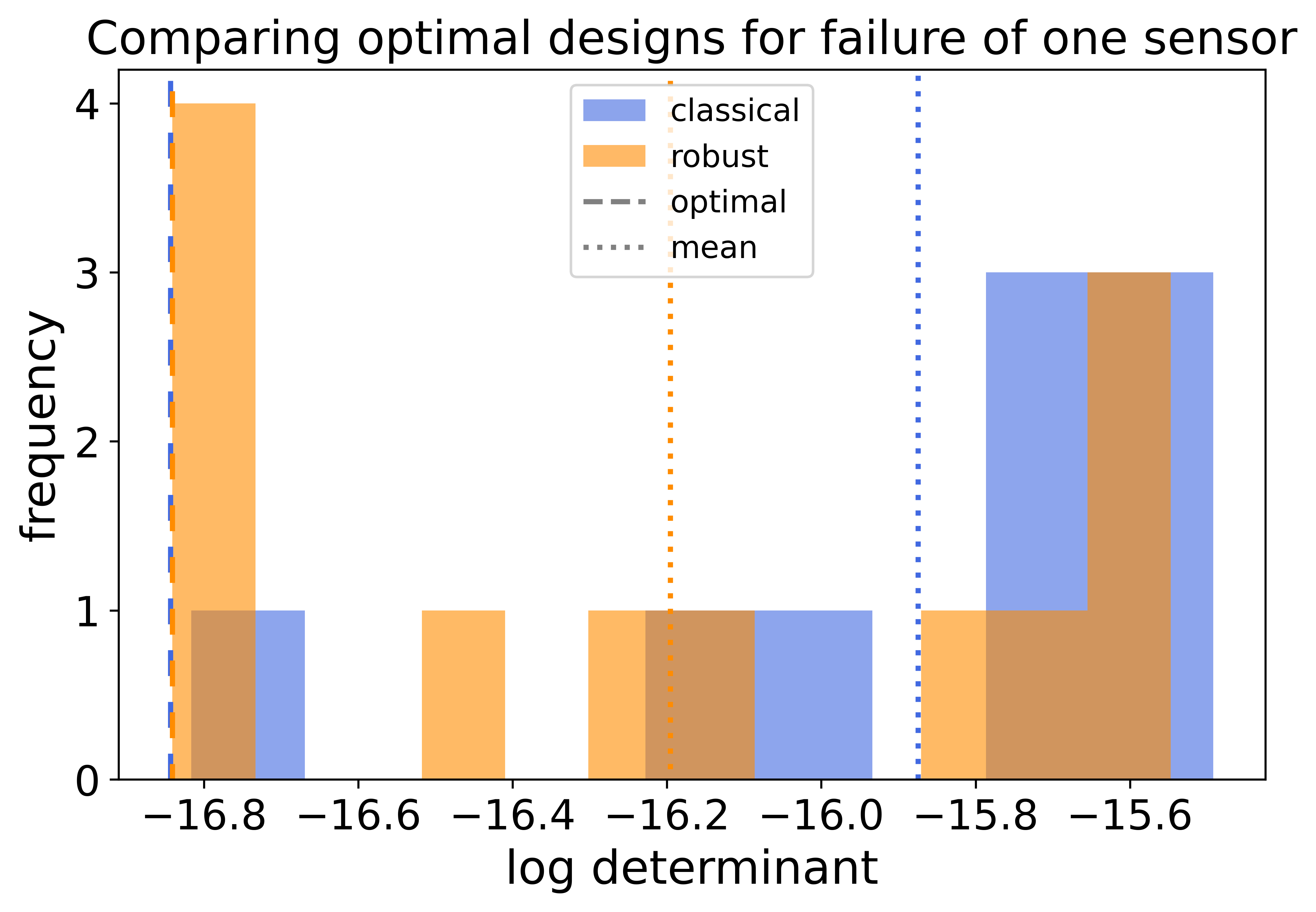} 
    \includegraphics[width=0.45\linewidth]{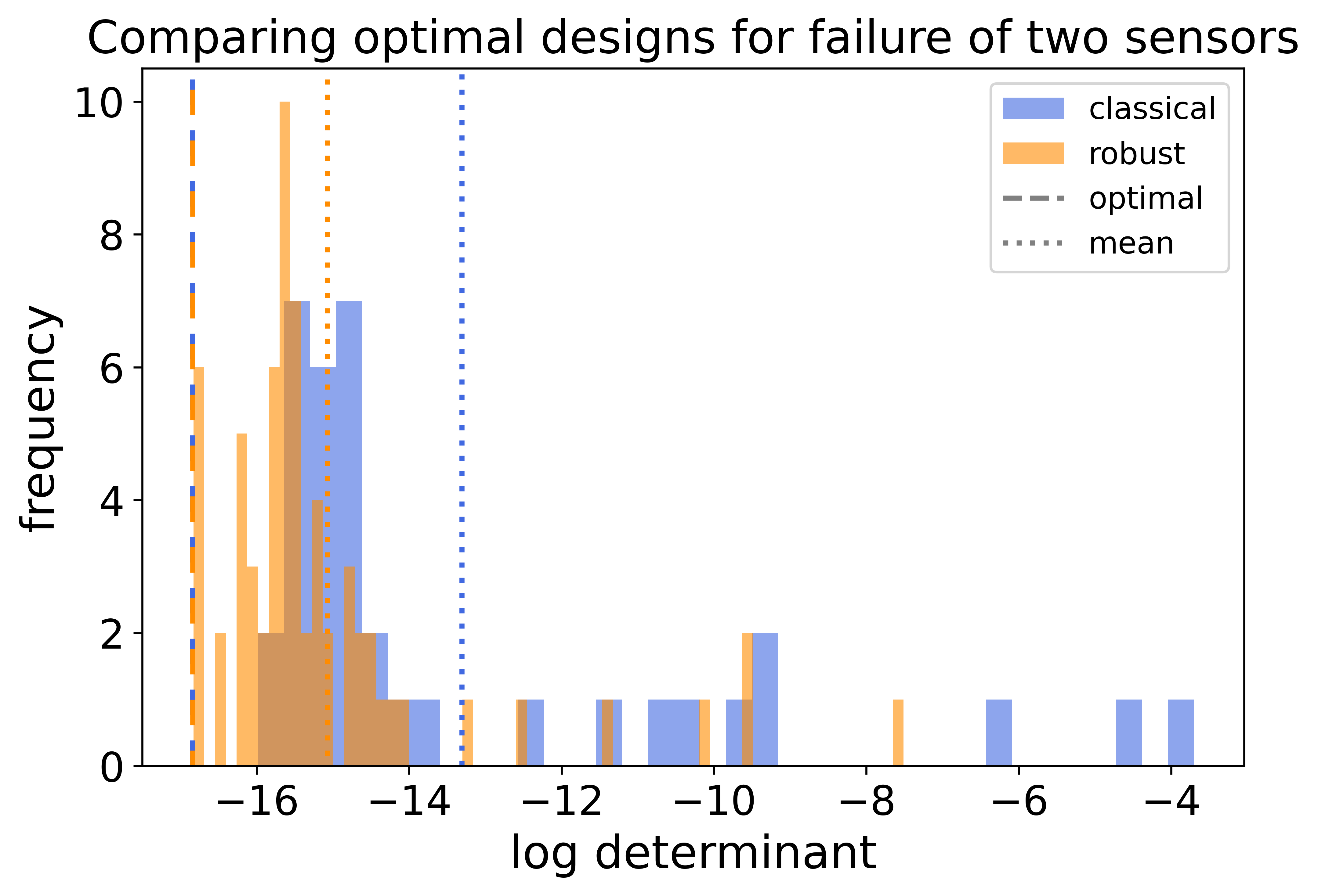} 
    \caption{Comparing the performance of the  robust versus classical fractional optimal designs. The histogram represents the performance over (left) one sensor or (right) two sensors failing, while the optimal (no sensors failing) and average performances are noted with the vertical lines.}
    \label{fig:AllDropOut_rob_v_class_opt_failure}
\end{figure}

As discussed in \Cref{sec:pp_fail_scen}, it can be informative to look at the behavior as one or more of the \textit{optimal} sensors fail to get a sense of how well the robust versus classical designs may perform in practice.
\Cref{fig:AllDropOut_rob_v_class_opt_failure} compares design performance in terms of the $\Psi := \log \det(\cdot)$ criterion when no sensors fail (noted as optimal) while providing a histogram of performances corresponding to one or two of the optimal sensors failing. 
The dotted line denotes the average performance over these realizations.
%
Note that in this context, accounting for one sensor failing amounts to zeroing out one of the fractional weights. 
For fairness of comparison, we then renormalize the designs before evaluating performance.
From the left plot of~\Cref{fig:AllDropOut_rob_v_class_opt_failure}, we see that although the robust and classical designs perform similarly when no sensor fails, the robust design, on average and in terms of the worst-case scenario, provides better information gain in the event that one of the optimal sensors fails. 
Since one cannot know if and which sensors will fail in a real experiment, it is ideal for the robust design to perform nearly as well as classical $D$-optimal design when no sensors fail, but to outperform the classical approach in the event of failure.

As noted in~\Cref{sec:pp_fail_scen}, solving the OED problem while considering the failure of $n_f >1$ sensors is computationally challenging due to the combinatorially many scenarios that must be considered. 
Hence, it is advantageous to examine in postprocessing if robust-to-one-sensor designs are, in fact, robust to multiple sensors failing.
From the right plot of~\Cref{fig:AllDropOut_rob_v_class_opt_failure}, we see that the robust design on average, and in terms of the worst-case scenario, outperforms the classical design in the event that two optimal sensors fail. 
Furthermore, when two sensors fail, the relative (to optimal) average information gain for robust versus classical designs differs more significantly than with a single sensor failing, indicating that the robustness of classical designs may diminish with increased sensor failures.
Although such a result is problem specific, it indicates that solving the more computationally tractable problem of designing against one sensor failure could provide  robustness to multiple sensors failing.

\subsection{Robustness to known probabilities of failure}\label{sec:comp_res_known_pof}

\begin{figure}[h!]
    \centering
    \includegraphics[width=0.9\linewidth]{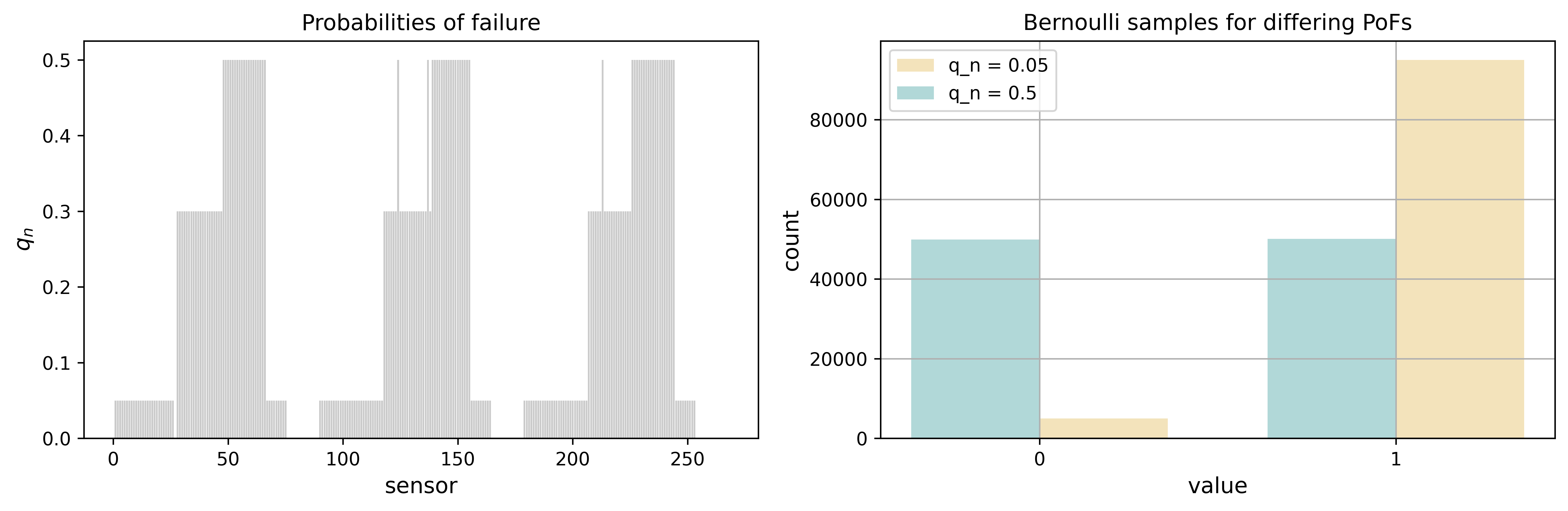} 
    \caption{(Left) a depiction of the probabilities of failure (PoFs) for each of the candidate sensor locations. (Right) a histogram of $10^{5}$ realizations of a Bernoulli random, where the color of the bar corresponds to the PoF.}
    \label{fig:PoF_bern_samps}
\end{figure}

In many engineering applications, one may have sufficient knowledge regarding what sensor locations are most likely to fail and would like to incorporate such probabilities of failure (PoFs) when determining an optimal experimental design strategy.
We first consider an example using the wedding cake model (see~\Cref{sec:model_prob}), where probabilities of failure for each candidate sensor location are directly specified.
Here, physical intuition and experience indicate that sensors towards the top of the wedding cake structure experience larger forces and thus have higher probabilities of failure. 
With this, we define the PoFs such that sensors at the base and level $1$ are assigned $q_n = 0.05$, sensors at level $2$ are assigned $q_n = 0.3$, and sensors at level $3$ are assigned $q_n = 0.5$.
A depiction of the PoFs for each of the $267$ candidate sensors is given in the left plot of~\Cref{fig:PoF_bern_samps}.

\begin{figure}[h!]
    \centering
    \includegraphics[width=0.7\linewidth]{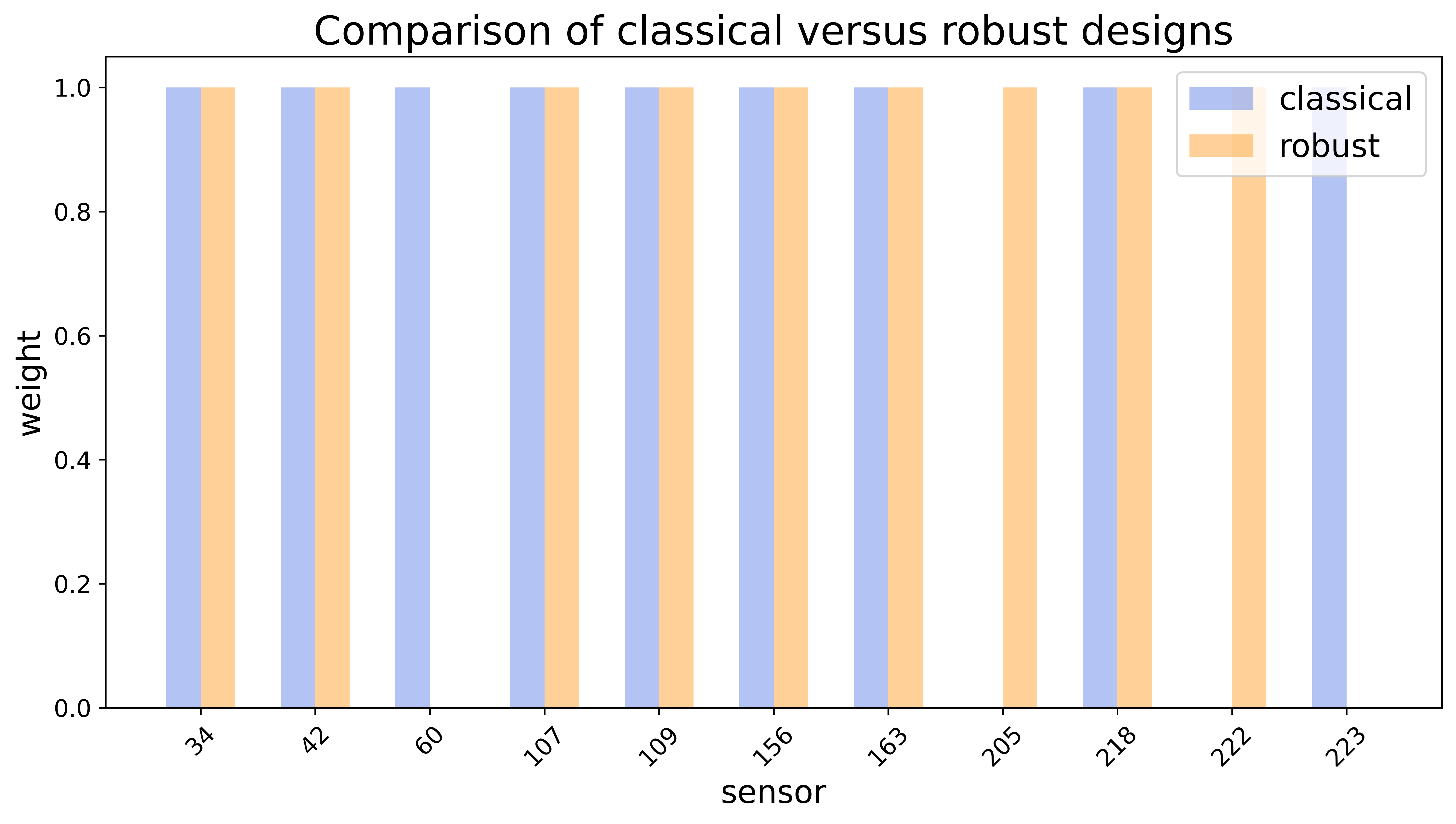}
    \caption{The binary optimal experimental designs associated with optimizing the classical OED criterion in~\eqref{eq:class_opt_des} and robust criterion in~\eqref{eq:rob_opt_des}, leveraging the binary-inducing double-well penalty.}
    \label{fig:PoF_rob_v_class_frac_des}
\end{figure}

\Cref{fig:PoF_rob_v_class_frac_des} compares the classical D-optimal design computed by optimizing~\eqref{eq:class_opt_des} versus the PoF formulation determined via~\eqref{eq:rob_opt_des}. 
Note that binary designs are not guaranteed for every value of $\gamma$ in~\eqref{eq:rob_opt_des} as this simply weights the importance of the OED criterion to the binary structure.
Hence, to ensure binary designs, we sweep over $100$ logarithmically spaced values of $\gamma$, ranging from $10^{-1}$ to $10^{5}$. 
We then select the optimal design as the binary solution that best minimizes the OED criterion of interest.
From~\Cref{fig:PoF_rob_v_class_frac_des}, we see that the resulting robust and classical designs share several optimal sensor locations but are distinct.
Notice that, unlike the approach given in~\Cref{sec:one_sensor_fails} (which requires sampling over failure scenarios), optimizing~\eqref{eq:rob_opt_des} is no more computationally expensive than the classical approach.

\begin{figure}[h!]
    \centering
    \includegraphics[width=0.45\linewidth]{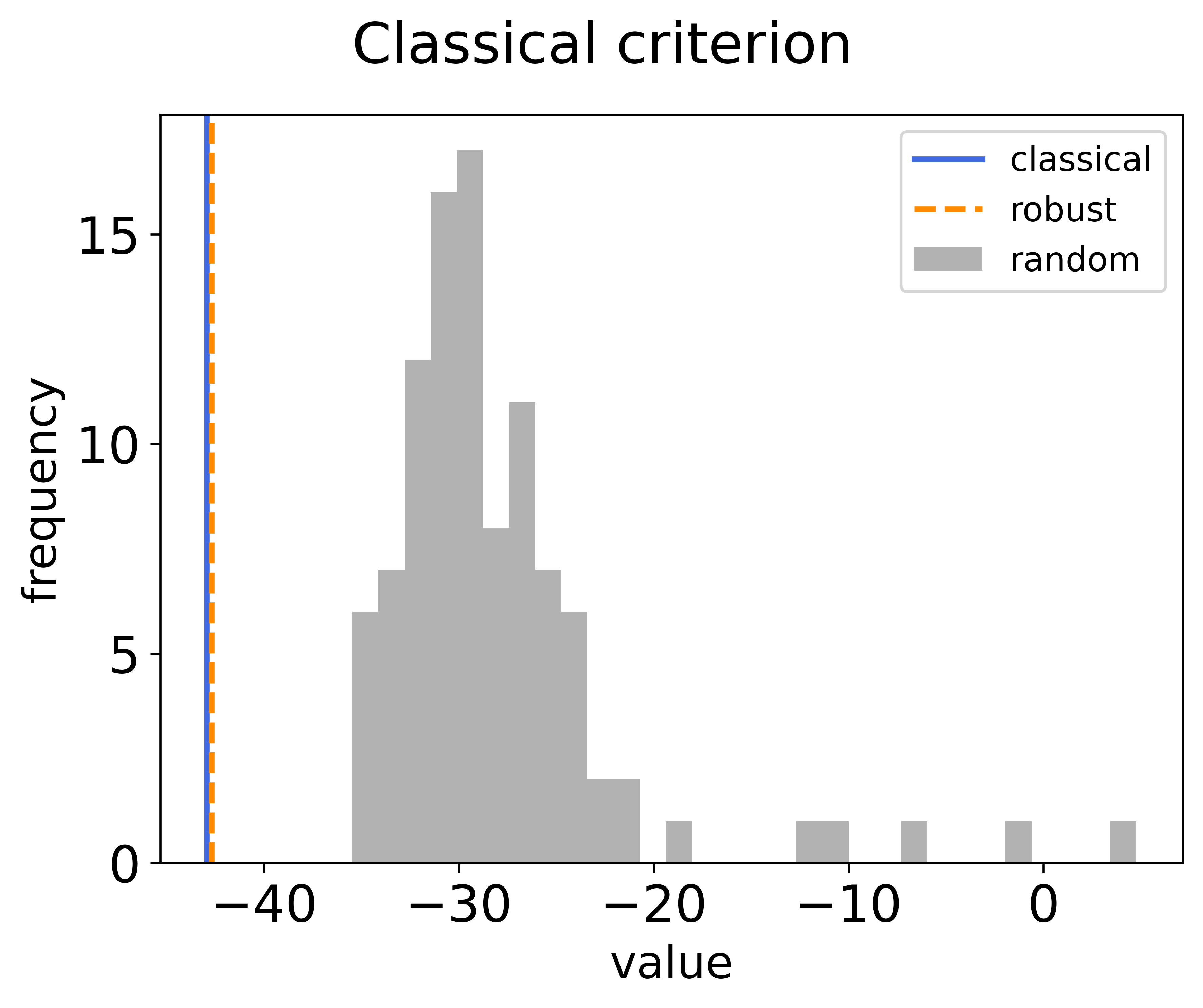} 
    \includegraphics[width=0.45\linewidth]{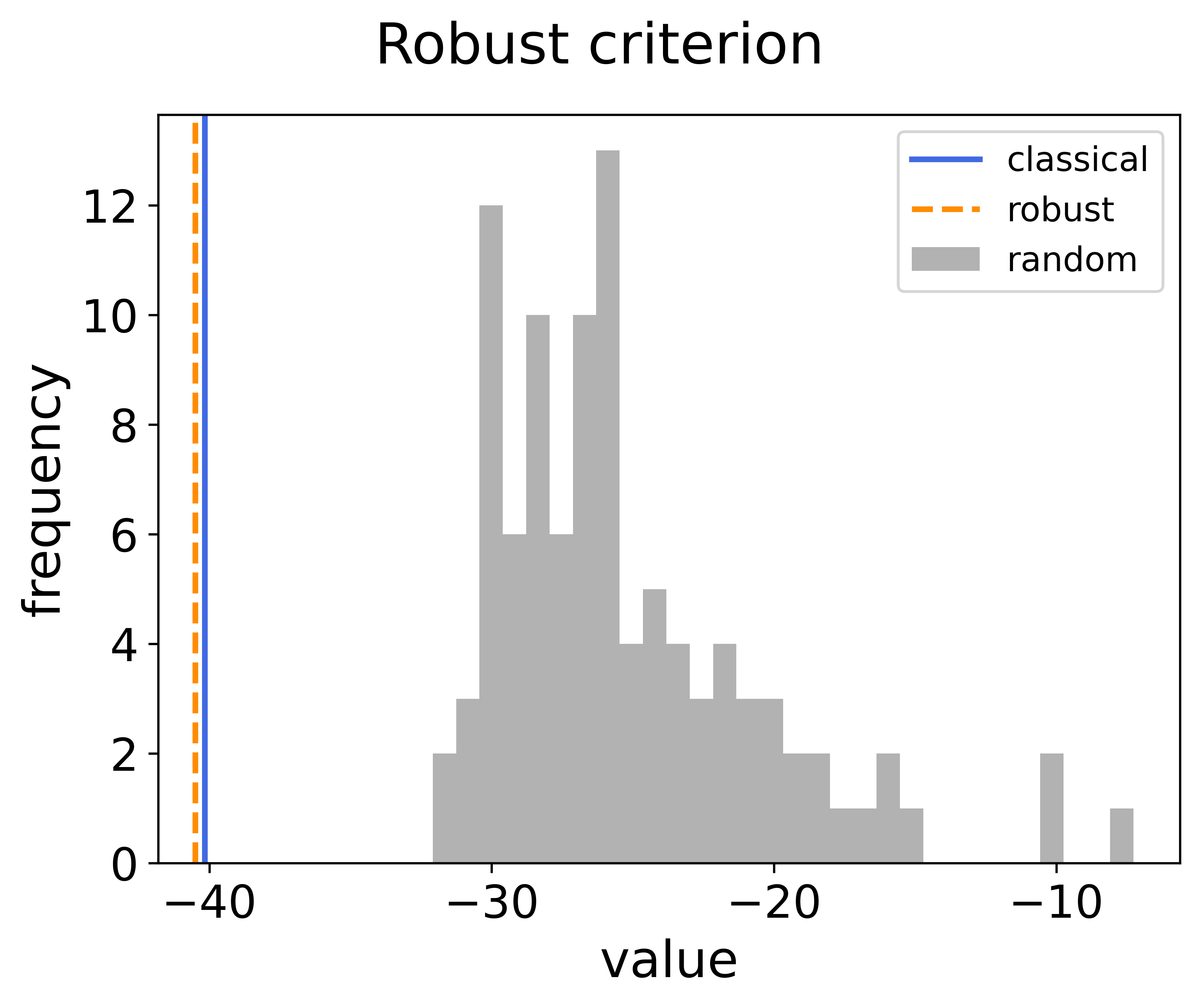}
    \caption{Comparing the classical and robust binary optimal designs versus random  designs based on the (left) classical OED criterion in~\eqref{eq:class_perf} versus the (right) robust criterion in~\eqref{eq:robust_perf}.}
    \label{fig:PoF_rob_v_class_frac_performance}
\end{figure}

\begin{figure}[h!]
    \centering
    \includegraphics[width=0.6\linewidth]{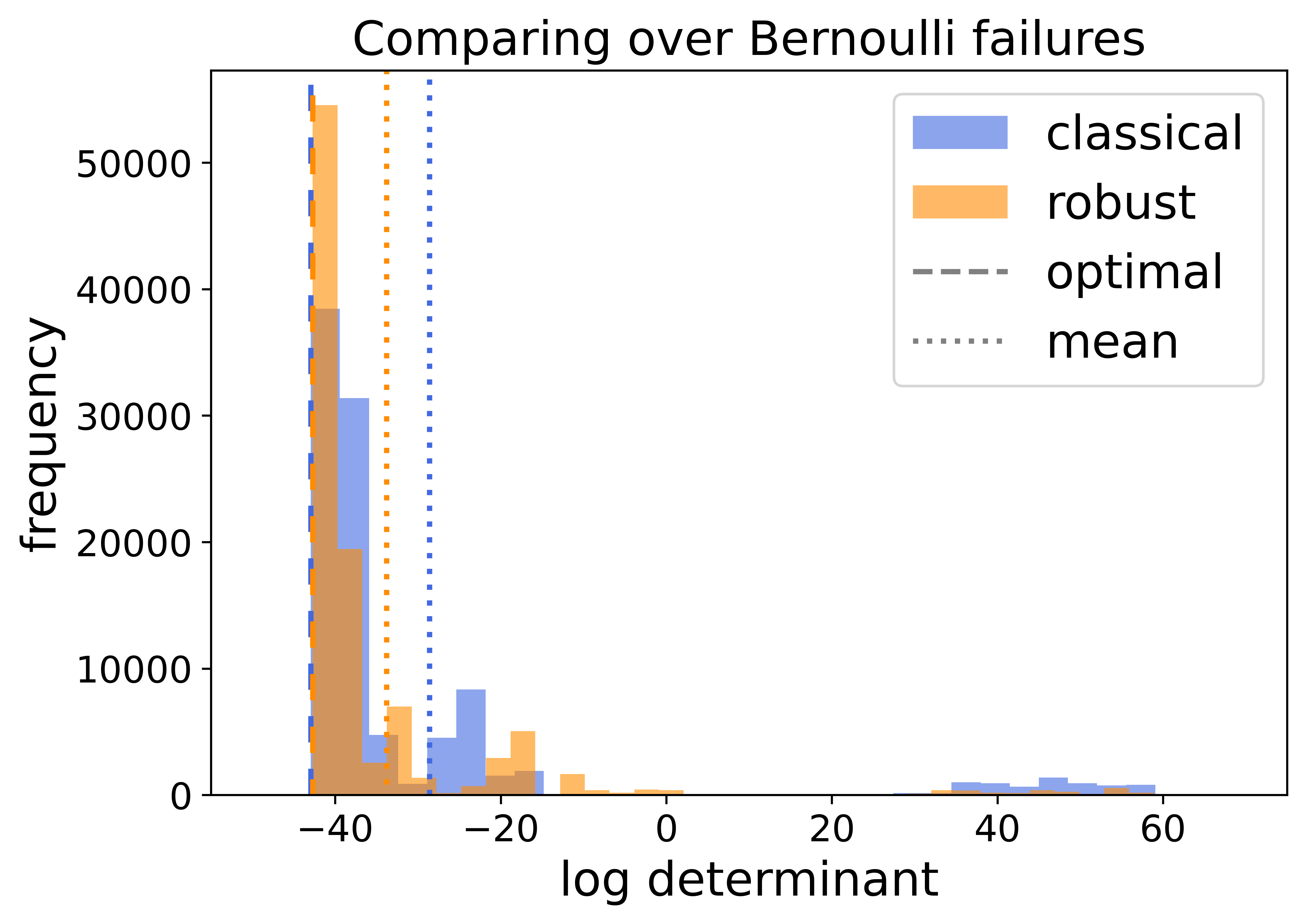} 
    \caption{Comparing the performance of the  robust versus classical binary optimal designs. The histogram represents the performance over failures determined by sampling Bernoulli random variables corresponding to the PoFs, where the histogram is depicted for log determinant values less than or equal to $75$. The optimal (no sensors failing) and average performances are noted with the vertical lines.}
\label{fig:PoF_rob_v_class_opt_bern_failure}
\end{figure}

In~\Cref{fig:PoF_rob_v_class_frac_performance}, we compare the performance in terms of the OED criteria used in optimization (see~\Cref{sec:post_proc}); 
this figure indicates that if no sensors fail, the robust and classical designs perform similarly in terms of their respective criteria, both outperforming random binary designs.
We next compare performance in terms of failure scenarios we could see in an experiment based on the assigned PoFs. 
To do so, we leverage the postprocessing metric described in~\Cref{sec:pp_fail_scen} for $j = 1, \dots, n_{\text{samps}}$ Bernoulli random variable samples, where $n_{\text{samps}}=10^5$.
\Cref{fig:PoF_rob_v_class_opt_bern_failure} provides a histogram of the performance of robust versus classical designs over the Bernoulli-sampled failure scenarios.
Because~\Cref{fig:PoF_rob_v_class_opt_bern_failure} depicts performance over all sampled failure scenarios, rather than just failure scenarios affecting the optimal sensors, it reflects both how likely an optimal sensor is to fail (based on the assigned PoFs) as well as the impact of such failures.
Since one sample of a Bernoulli random variable can result in many sensors failing for a given experiment, and hence very large log-determinant values, we omit from the histogram performances greater than $75$ for ease of visualization. 
Such poor performances are, however, accounted for in the mean (dotted line).
The result depicted in~\Cref{fig:PoF_rob_v_class_opt_bern_failure} indicates that by explicitly incorporating knowledge regarding how likely sensors are to fail, one can increases robustness of the sensor placement strategy to such failures, with no additional computational cost.

Furthermore, we note that the postprocessing metric (see~\Cref{sec:pp_fail_scen}) is similar to alternative formulations of optimization objectives that average optimality criterion over realizations of Bernoulli random variables; see~\cite{smucker_robustness_2017}.
However, in practice, such criteria may be unstable if the sampling results in too many sensors failing and are much more expensive to compute (rather than one criterion evaluation, you compute $n_{\text{samps}}$ evaluations, where $n_{\text{samps}}$ is the number of random variable samples)\footnote{Note that for this example, none of the samples resulted in the inverse problem becoming ill-posed, and thus all were considered in postprocessing.}.
Thus, the result given in~\Cref{fig:PoF_rob_v_class_opt_bern_failure} indicates that the more computationally tractable objective provided by~\eqref{eq:rob_opt_des}, may provide the desired behavior.
That is, robust designs may be, on average (over failure scenarios), more $D$-optimal than classical designs.

\begin{figure}[h!]
    \centering
    \includegraphics[width=0.85\linewidth]{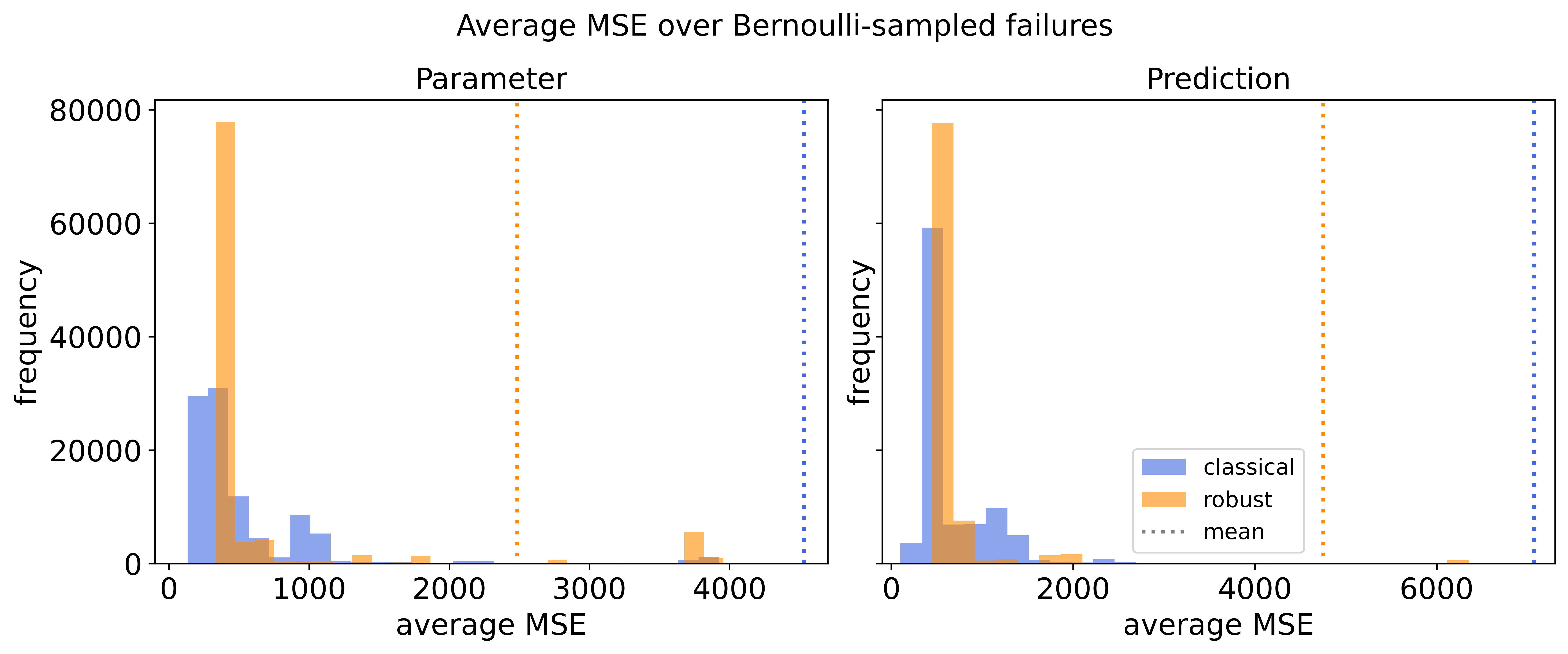} 
    \caption{A comparison of average parameter (left) and prediction (right) mean squared errors (MSEs) for robust versus classical designs over failure scenarios determined by sampling Bernoulli random variables. The dotted line indicates average (over failure scenarios) performance (average MSE).}
    \label{fig:param_pred_mse_fail}
\end{figure}

As noted in~\Cref{sec:mse}, comparison in terms of classical $D$-optimal criterion (log-determinant) does not provide a direct measure of parameter and prediction mean squared errors (MSEs). 
As such, we compare classical versus robust designs in terms of MSE across realizations of nominal parameter values. 
Here, we draw $n=100$ samples from a multivariate Gaussian distribution with a mean given as $\mu_{\theta_0} = \left[4\mathrm{e}01, \hspace{0.1cm} 5\mathrm{e}02, \hspace{0.1cm} 2\mathrm{e}00, \hspace{0.1cm} 1.2\mathrm{e}01, \hspace{0.1cm}6\mathrm{e}02, \hspace{0.1cm}8\mathrm{e}01\right]^\top$ and covariance given by $6^2{\bm I}$, where ${\bm I} \in\mathbb{R}^{\paramdim \times \paramdim}$.
To understand how MSE performance varies over failure scenarios, we again leverage Bernoulli random variables to represent failure scenarios aligned with the assigned sensor PoFs.
Here, we draw $n_{\text{samp}} = 10^5$ Bernoulli samples and compare the average MSEs over the failure scenarios.
\Cref{fig:param_pred_mse_fail} depicts the performance of the robust and classical designs, where we see that over a range of failure scenarios, the robust design outperforms classical in terms of parameter and prediction MSE.
Again, for visualization we remove performance outliers beyond the $x$-axis limits, while accounting for these in the sample mean.

Note that in~\Cref{fig:param_pred_mse_fail} we plot all Bernoulli-sampled failure scenarios, not just those affecting the optimal design.
Hence, this result indicates that based on the likelihood of sensor failures and the performance of designs when sensors fail, the robust design will (on average) outperform the classical. 
Of course, for some failure scenarios, the classical design could outperform the robust in terms of average MSEs.
Furthermore, robust and classical optimal designs have nearly identical median performance, with the median average parameter MSE being $\approx 350$ and median average prediction MSE being $\approx 480$, which could indicate that the robust design is avoiding rare, but potentially very poor performances.
In~\Cref{fig:param_pred_mse_fail}, one can see that if no sensors fail, the classical design outperforms the robust in terms of parameter and prediction average MSE (as noted by the minimum value associated with classical designs being less than the minimum value associated with robust designs).
However, without \textit{a-priori} knowledge of what will happen in an actual experiment, one relies on probabilistic assessments of design performance.


\subsection{Dropout due to sensor clipping}\label{sec:clipping}

\begin{figure}[h!]
    \centering
    \includegraphics[width=0.8\linewidth]{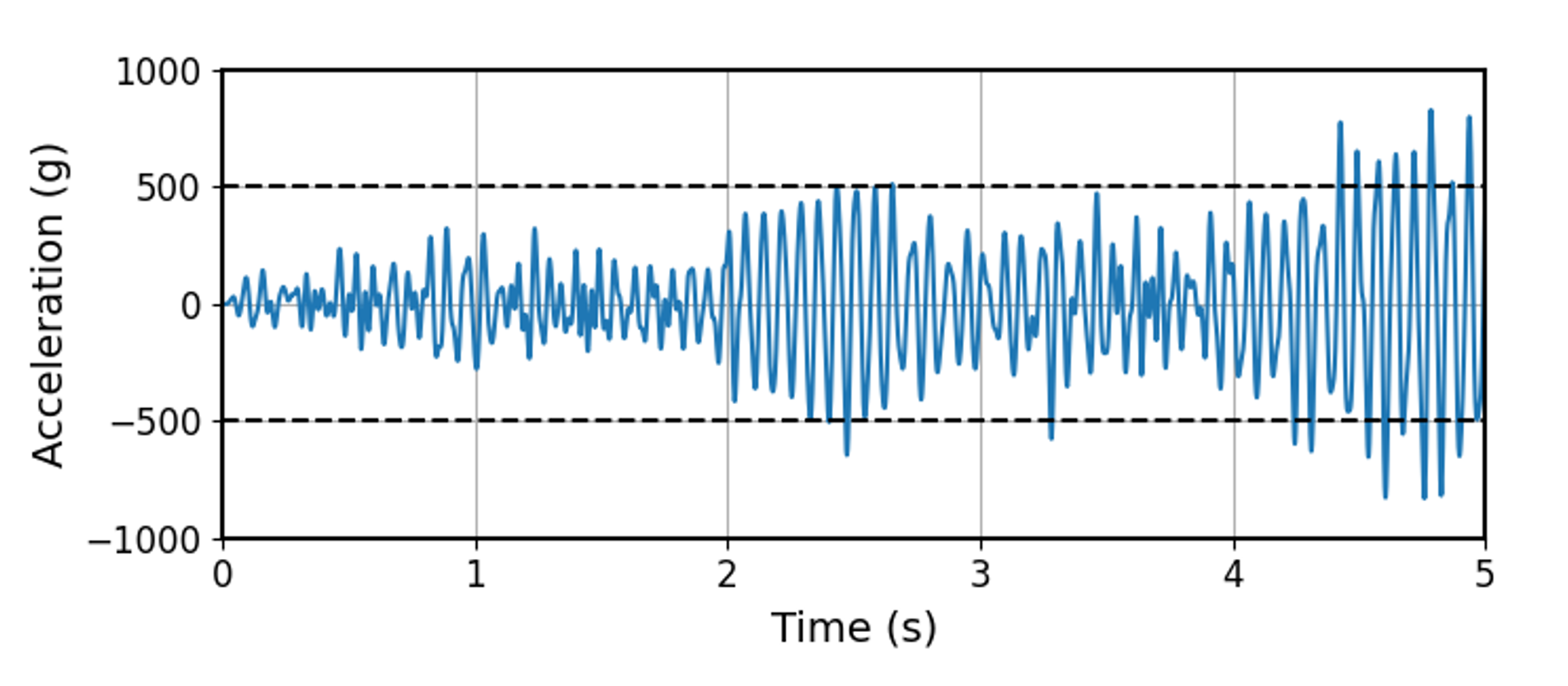}
    \caption{The acceleration response of a candidate sensor that exhibited sensor clipping. The clipping threshold is shown as dashed lines.}
    \label{fig:output_snapshots}
\end{figure}

Here, we provide computational results demonstrating a potential application of our robust OED formulation to mitigate the loss of critical information caused by sensor clipping in high-acceleration environments. 
Sensor clipping occurs when the measured vibration exceeds the dynamic range of the accelerometer, leading to distortion or complete loss of data. 
One can account for such clipping when designing sensor placements by first generating 
dropout scenarios that are incorporated into the robust OED framework.
Given prior knowledge of the loading conditions in the form a probability distribution, we can sample and subsequently predict the response at each candidate sensor location.
Dropout samples are then constructed by determining if the predicted response exceeded the specified clipping threshold.
Figure~\ref{fig:output_snapshots} presents an example where the predicted response at a single candidate sensor is clipped as the time-domain force exceeds the threshold noted by the horizontal dashed line.

\begin{figure}[h!]
    \centering
    \includegraphics[width=1.0\linewidth]{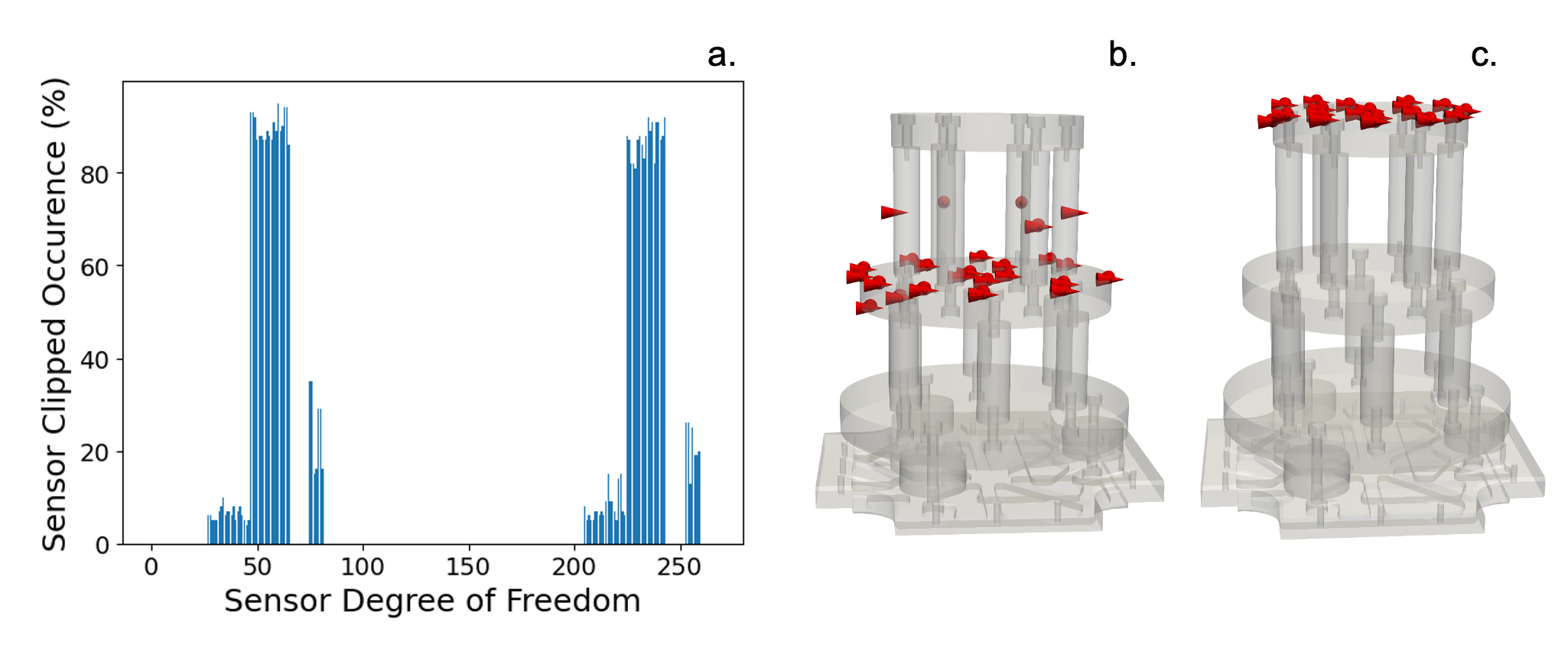}
    \caption{a) Percentage occurrence of dropout for each candidate sensor. b) Location of sensors that dropped out between $5\%\textrm{ and }20\%$  of the time and c) those that dropped out over 80\% of the time.}
    \label{fig:dropout-clipp-samples}
\end{figure}

\begin{figure}[h!]
    \centering
    \includegraphics[width=0.7\linewidth]{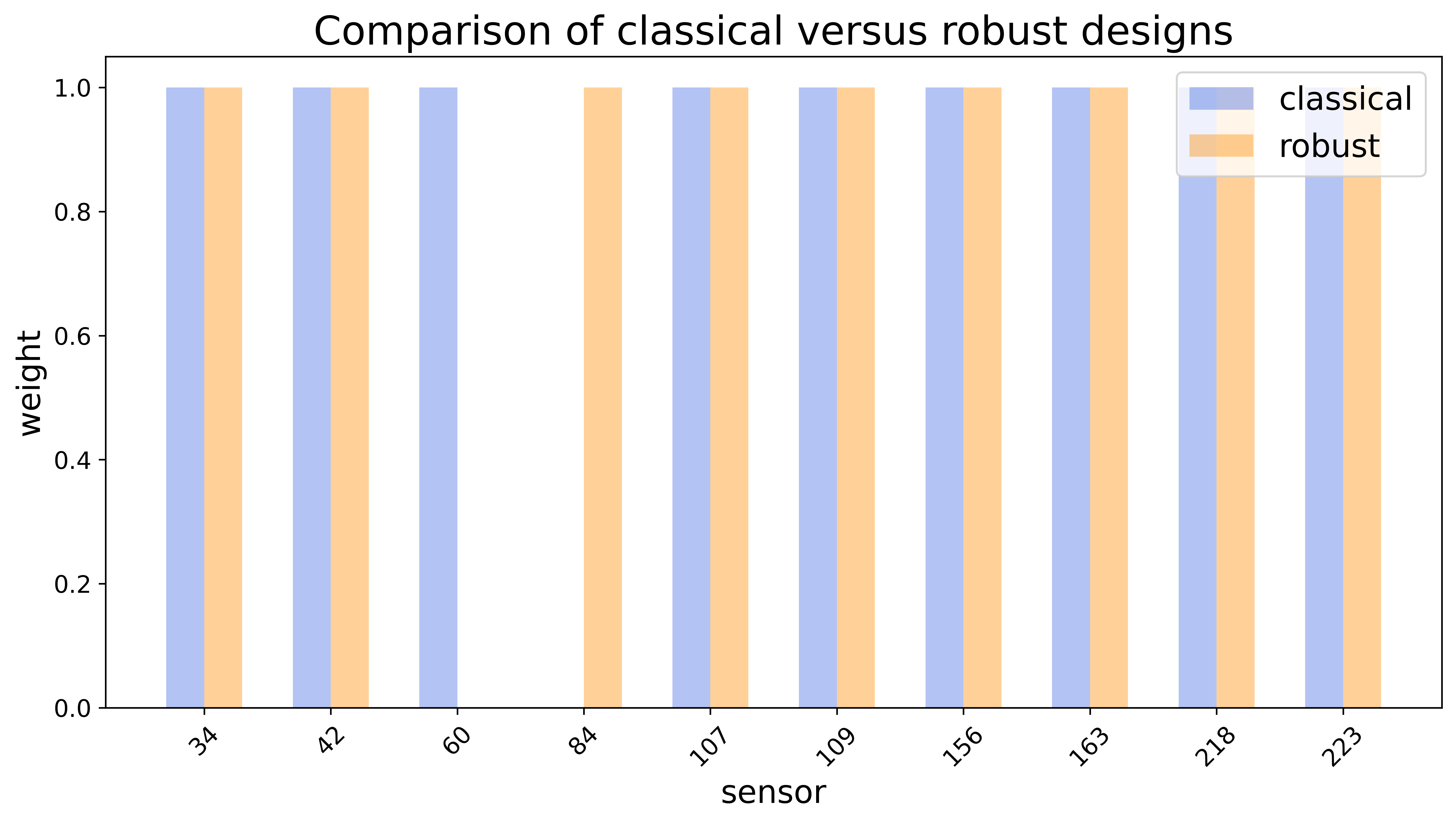}
    \\
    \begin{minipage}{0.45\linewidth}
    \centering
    \includegraphics[scale=0.18]{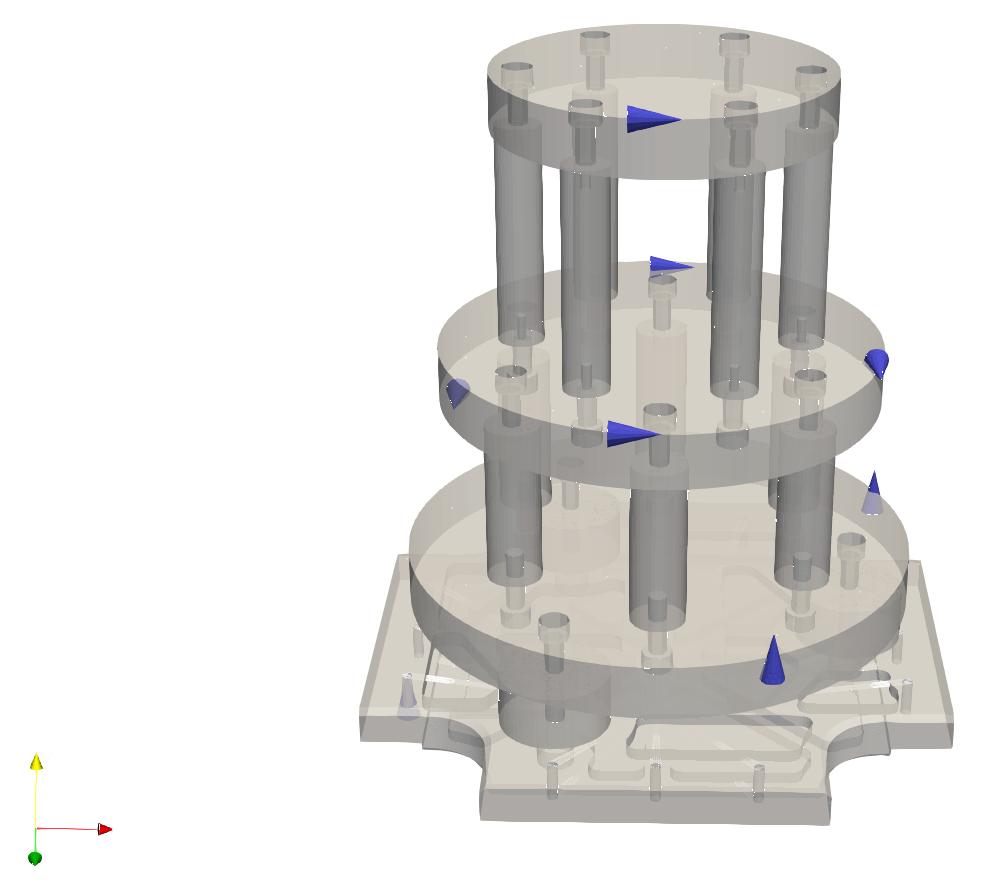}
    \caption*{(a) Classical optimal design}
    \end{minipage}
      \hfill
    \begin{minipage}{0.45\linewidth}
    \centering
     \includegraphics[scale=0.16]{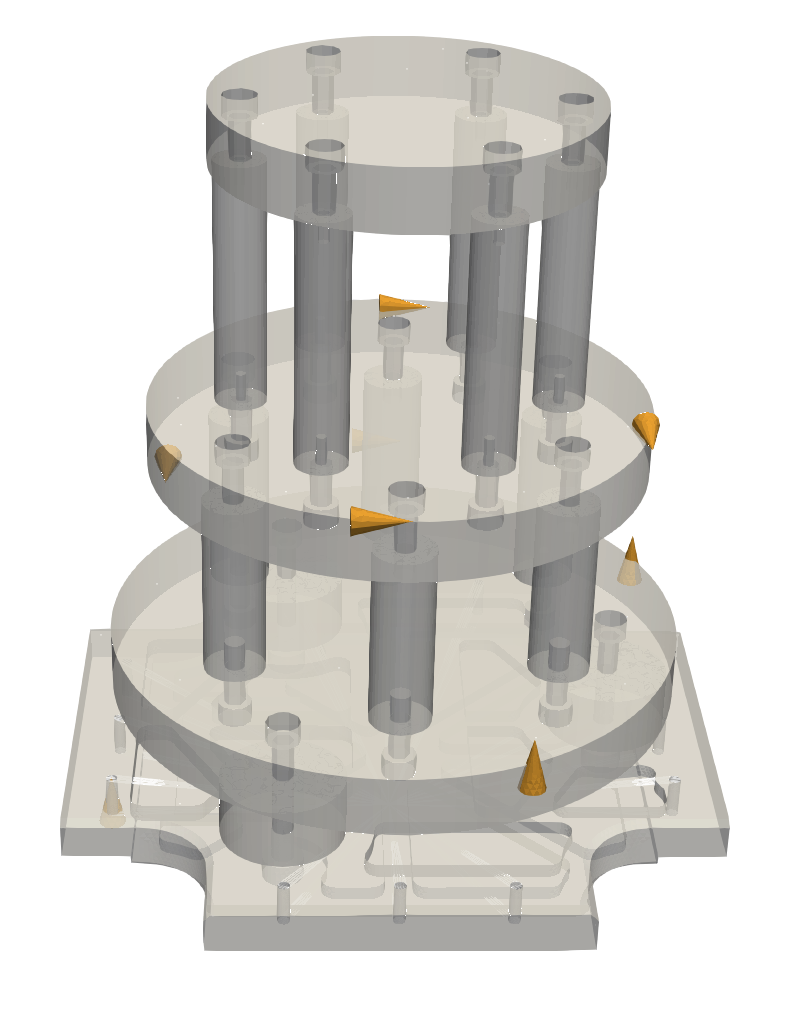}
    \caption*{(b) Robust optimal design}
    \end{minipage}
    \caption{(Top) The binary optimal experimental designs associated with optimizing the classical OED criterion in~\eqref{eq:class_opt_des} and robust criterion in~\eqref{eq:rob_opt_des}, leveraging the binary-inducing double-well penalty.
    (Bottom) The corresponding optimal classical (a) and robust (b) optimal designs (physical location and orientations) imposed on the wedding cake structure.}
    \label{fig:binary_clipped_rob}
\end{figure}

To illustrate this approach, we employ the wedding cake model (\Cref{sec:model_prob}) and simulate a scenario where the accelerometer clipping threshold is set to $500$ $g$'s. 
The forces acting at the base are modeled as an uncorrelated multivariate Gaussian distribution with zero mean. We specify the variance of the Gaussian large enough to ensure that clipping occurred at sensor locations with significant vibration response, thereby illustrating the effects of clipping on the design of optimal sensors.
One hundred time-domain force realizations are generated, and the structural response at all candidate sensor locations is computed using a finite element model. 
\Cref{fig:dropout-clipp-samples} depicts the dropout scenarios generated. 
Plot (a) depicts the percentage occurrence (of the $100$ force realizations) of dropout for each of the candidate sensor locations, while the right plot shows the physical location of sensors that fail $20\%$ (b) versus $80\%$ (c) of the time.
Intuitively, level $3$ of the wedding cake structure experiences the largest accelerations, and hence has the highest rate of clipping.
Level $2$ experiences smaller accelerations resulting in the sensors that fail between $5\%\textrm{ and }20\%$ of the time; see~\Cref{fig:candidate_sensor_fem} for reference. 
The clipped occurrence frequencies are incorporated into the robust OED optimization problem as follows.
If candidate accelerometer sensor $i$ is predicted to exceed the clipping threshold, it is noted as a dropout sample with $\{\pmat_j\}_{ii} = 0$ in~\eqref{eq:mult_rob_opt_des}, for the $j=1,\dots, 100$ force realizations considered.
With these defined dropout scenarios, we then solve the robust OED problem defined in~\eqref{eq:mult_rob_opt_des} and compare the results to classical OED.
\Cref{fig:binary_clipped_rob} depicts the binary optimal designs (top row) and the corresponding designs imposed on the physical wedding cake structure.
Notice from~\Cref{fig:binary_clipped_rob}, the robust design avoids placing sensors at the top plate, which is intuitive as such sensors have the largest predicted acceleration responses, and hence clipping is likely to occur. 

\begin{figure}[h!]
    \centering
    \includegraphics[width=0.6\linewidth]{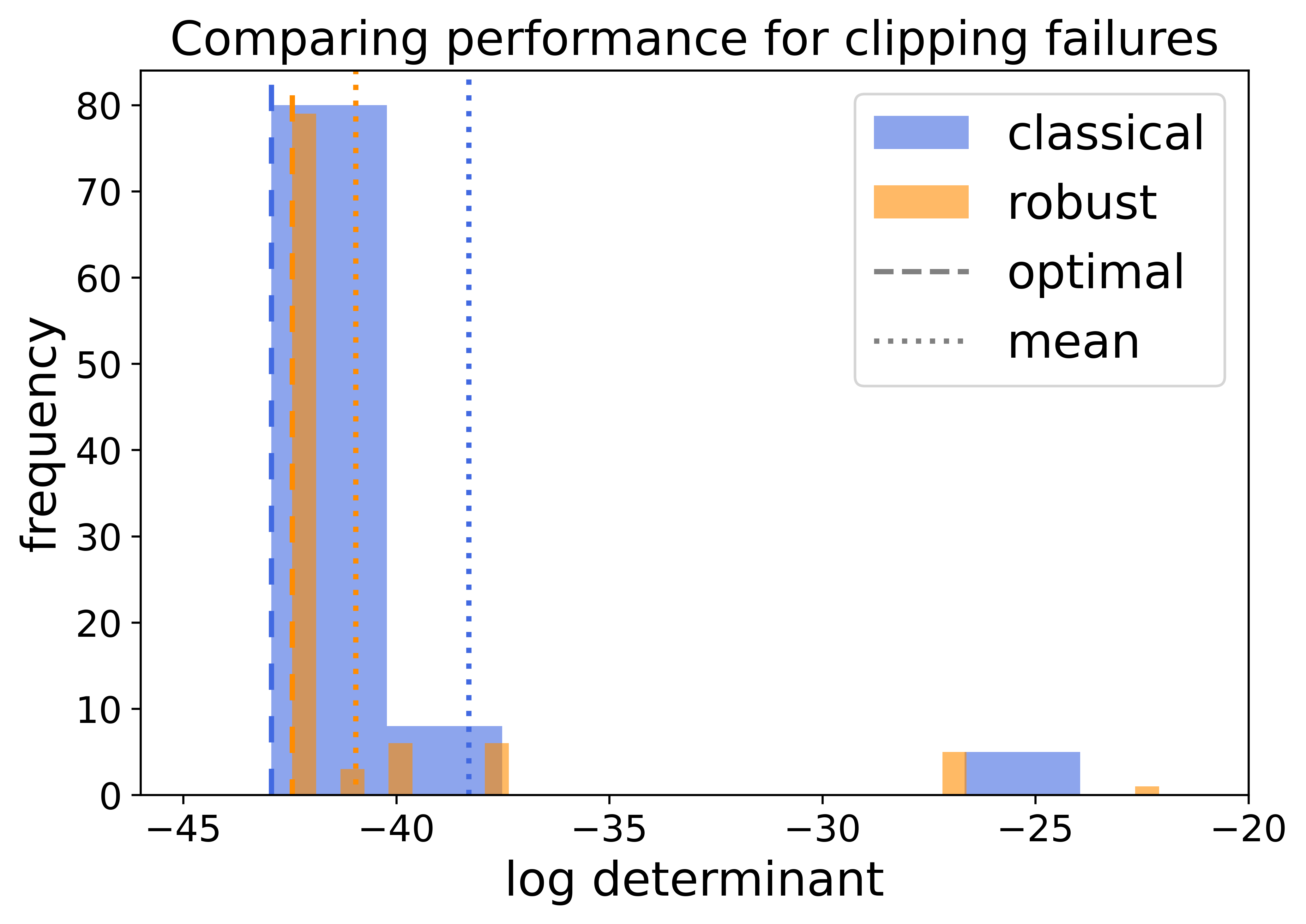}
    \caption{Comparing the performance of the  robust versus classical binary optimal designs. The histogram represents the performance over failures determined by predicting clipping behavior over $j=100$ random initial force realizations. The histogram is depicted for log determinant values less than or equal to $-20$. The optimal (no sensors failing) and average performances are noted with the vertical lines.}
    \label{fig:clipped_perf}
\end{figure}

We compare the performance of the two designs depicted in~\Cref{fig:binary_clipped_rob} by looking both at how well the designs perform with no failures (i.e. using the optimality criteria given in~\eqref{eq:class_perf} and~\eqref{eq:robust_perf}) as well as looking at the behavior when failures occur according to the predicted $\{\pmat_j\}_{1}^{100}$ (see~\Cref{sec:pp_fail_scen}).
\Cref{fig:clipped_perf} provides a histogram of performance values over the $j=100$ clipping scenarios, with the vertical lines representing the optimal (dashed) and mean (dotted) performance.
For visualization, we restrict the $x$-axis to performance values less than or equal to $-20$. however, such performances are accounted for by the mean.
Since the clipped occurrence percentage is sampled from random initial forces, we compare performance for all $j=100$ samples---even if a sample does not correspond to an optimal sensor being dropped. 
Thus, the result provides a sense of both 1) the likelihood of an optimal sensor being clipped and 2) the impact of clipping on both designs. 

\FloatBarrier
\section{Conclusions}\label{sec:conclusions}
This works explores the application of optimal experimental design (OED) formulations that are robust to sensor dropout (failure) in the context of a finite-element model of a three tiered, wedding cake structure.
Here, accelerometer data is used to inform the uncertain model parameters given by the nonzero loads applied to the base of the structure.
We compare classical OED approaches to robust formulations aimed at maintaining performance in the event of sensor failures.
Specifically, we consider two scenarios relevant to these application problems: 1) when the probabilities of failure of each candidate sensor are known or can be predicted and 2) where failure probabilities are not known, so all failure scenarios of interest must be accounted for directly.
We also demonstrate how such robust formulations can be applied to clipping scenarios, where measurements are corrupted when measured vibrations exceed the dynamic range of the sensor.

Computationally, we leverage a relaxation-based approach that avoids the combinatorially complexity of optimizing binary designs. 
Here, the designs are casts as fractional weights enabling efficient gradient-based optimization,
where a novel \textit{double-well} penalty induces a binary structure on the designs.
Such an approach provides a way to determine binary designs during optimization as opposed to post-optimization heuristic approaches, such as rounding. 
To compare the performance of robust versus classical $D$-optimal designs, we evaluate performance in terms of the optimization criterion as well as parameter and prediction mean-squared errors. 
We also evaluate these performances over the relevant failure scenarios, i.e. with respect to optimal sensors failing, to understand potential experimental outcomes.

We find that when probabilities of failure can be specified for each candidate sensor, accounting for such knowledge of failures results in  robust designs that differ from classical designs.
We evaluate performance of the designs by sampling $10^5$ plausible experimental scenarios, determined by sampling Bernoulli random variables whose parameters correspond to the specified probabilities of failure.
We find that if no sensors fail, robust and classical designs perform similarly in terms of the log-determinant. 
Furthermore, the robust design outperforms the classical in terms of the average (over failure scenarios) log-determinant of the parameter covariance as well as prediction and parameter mean squared errors.
We note that in comparison to alternative formulations~\cite{smucker_robustness_2017}, our approach allows for varying probabilities of failure over the candidate design space and provides robust designs for the same computational cost as computing classical designs.

When probabilities of failure cannot be specified, we consider OED formulations that average over specified dropout scenarios. 
The optimal design is then determined as one with the best (on average) performance.
Note that in such problems, computing the robust design is $k$ times more expensive than computing the classical design, where $k$ is the number of failure scenarios of interest.
Here, we consider fractional designs
as the binary counterparts were the same for robust and classical designs. 
While the framework is general, we formulate the OED optimization problem to be robust to the dropping of one sensor.
Performance is then evaluated in terms of the log-determinant of the parameter covariance, and we consider how performance varies as one of the optimal sensors is dropped.
We find that if no sensors fail, the robust and classical designs perform similarly; however, the robust design outperforms the classical (on average) as one sensor is lost. 
Furthermore, we find that robust outperforms classical when considering the failure of two optimal sensors, indicating that one may be able to solve the more tractable problem of guarding against one sensor failure, while maintaining additional robustness.

Overall, for the model and design problems of interest, we find that the classical and robust designs share many of the same optimal sensor locations; however, over when evaluating performance over various failure scenarios of interest, robust designs outperform classical on average. 
Although the degree to which robust designs outperform classical could vary with the model and design problems considered, the similarity in the optimal designs may indicate that $D$-optimality is naturally robust to some failure scenarios. 
Future work could explore extensions of the robust approaches to multi-objective design problems as well as approaches to enabling tractable computation of the robust objectives for expensive, nonlinear models.

\section{Acknowledgments}\label{sec:ack}

This work was supported by the Advanced Simulation and Computing (ASC) Verification \& Validation (V\&V) portfolio at Sandia National Laboratories, a multimission laboratory managed and
operated by National Technology and Engineering Solutions of Sandia LLC, a wholly owned
subsidiary of Honeywell International Inc. for the U.S. Department of Energy’s National Nuclear
Security Administration under contract DE-NA0003525.
This article has been authored by an employee of National Technology \& Engineering Solutions of Sandia, LLC. The employee owns all right, title and interest in and to the article and is solely responsible for its contents. The United States Government retains and the publisher, by accepting the article for publication, acknowledges that the United States Government retains a non-exclusive, paid-up, irrevocable, world-wide license to publish or reproduce the published form of this article or allow others to do so, for United States Government purposes. The DOE will provide public access to these results of federally sponsored research in accordance with the DOE Public Access Plan \url{https://www.energy.gov/downloads/doe-public-access-plan}.

\section*{Appendix}\label{sec:app}

The covariance given in~\eqref{eq:do_cov} leveraged asymptotic arguments to account for probabilities of failure.
However, if ${\bm \Gamma}_{\text{noise}}=\sigma^2{\bm I}$, the non-asymptotic covariance can be derived as
\begin{eqnarray}\label{eq:do_non_asym_cov}
    {\bm C}({\bm\Xi}\des) = \sigma^2 \mathbb{E}[(\transmat^\top{\bm W}{\bm \Xi}\transmat)^{-1}]
  \approx \frac{\sigma^2}{N}\sum_{n=1}^N (\transmat^\top{\bm W}{\bm \Xi}_n\transmat)^{-1}
  =\frac{1}{N}\sum_{n=1}^N{\bm C}({\bm\Xi}_n\des),
\end{eqnarray}
where ${\bm \Xi}$ denotes the diagonal matrix with $\Xi_i$ on the diagonal, and ${\bm\Xi}_n$ denotes the $n$-th sample of ${\bm\Xi}$. 
Such a formulation is similar in spirit to that which is considered in~\cite{smucker_robustness_2017}. 
In comparison to our robust criterion defined in~\eqref{eq:rob_opt_des} as $\Psi(\des) = \log \det\big({\bm C}(\pmat\des)\big)$,~\cite{smucker_robustness_2017} instead considers 
\begin{eqnarray}\label{eq:smucker_crit}
    \psi(\des) = \mathbb{E} \left[ \det({\bm T}^{\top}{\bm D}{\bm T})^{\frac{1}{\paramdim}} \right],
\end{eqnarray}
where 
\begin{eqnarray}\label{eq:bernoulli2}
    d^2_i = 
    \begin{cases}
        1, \quad \text{with probability} \quad 1-q({\bm x}) \\
        0, \quad \text{with probability} \quad q({\bm x}),
    \end{cases}
\end{eqnarray}
and the expectation in~\eqref{eq:smucker_crit} is with respect to realizations of $d_i$.
While authors provide the formulation for more general probabilities of failure that can vary with design choice in~\eqref{eq:bernoulli2}, for computational tractability, they restrict $q({\bm x}) = q$ to approximate the expectation as the following series expansion.  
\begin{eqnarray}\label{eq:HA_crit}
    \mathbb{E} \left[ \det({\bm T}^{\top}{\bm D}{\bm T})^{\frac{1}{\paramdim}} \right] &=& 
    (1-q)^n\det({\bm T}^{\top}{\bm T})^{\frac{1}{\paramdim}}
    + 
    q(1-q)^{n-1}\sum_{i=1}^{n} \det({\bm T}^{\top}_i{\bm T}_i) ^{\frac{1}{\paramdim}}
    \\
    &+& 
    q^2(1-q)^{n-2}\sum_{\substack{i\neq j \\ i < j}}\det({\bm T}^{\top}_{ij}{\bm T}_{ij})^{\frac{1}{\paramdim}} \\\label{eq:smucker_final}
    &+&
    q^3(1-q)^{n-3}\sum_{\substack{i\neq j\neq k \\ i < j <k}} \det({\bm T}^{\top}_{ijk}{\bm T}_{ijk})^{\frac{1}{\paramdim}}
    + \dots 
\end{eqnarray}
where $n$ is the number of observations, and ${\bm T}_{ij}$ refers to the matrix ${\bm T}$ where the $i^{\text{th}}$ and $j^{\text{th}}$ columns are removed. 

At a high level, our formulation given in~\eqref{eq:rob_opt_des} leverages asymptotics in the computation of the covariance operator. 
While this provides a computationally tractable OED criterion as well as flexibility in varying the probabilities of failure as a function of candidate sensor location, asymptotic arguments may not hold for finite data.
In comparison, the HA criterion in~\eqref{eq:smucker_final} makes a simplifying assumption regarding the probabilities of failure that ensures a closed-form series representation of the OED criterion. 
We note, however, that evaluating~\eqref{eq:smucker_final} requires up to $\sum_{i=0}^{n-\paramdim} {n \choose i}$ determinants, making it significantly more computationally expensive. 
Furthermore, when more than $n-\paramdim$ observations are missing, the inverse problem becomes ill-posed.
To avoid such issues and maintain computational tractability, authors consider a truncated version of~\eqref{eq:smucker_final}.
Furthermore, while the formulation given in~\eqref{eq:smucker_final} bears resemblance to the criterion given in~\eqref{eq:mult_rob_opt_des}, we note that the latter is used to assess robustness to \textit{specified} failure scenarios (i.e. failure of one sensor) not probabilities of failure; thus, it does not require sampling over the failure of $i=1, \dots, n$ failures, which avoids the aforementioned computational issues.

\section{References}
\bibliographystyle{unsrturl}
\renewcommand{\refname}{}
\bibliography{robust_OED}

\end{document}